\documentclass[preprint]{aastex}

\newcommand{\arcm}{\ifmmode {' }\else $' $\fi}
\newcommand{\arcs}{\ifmmode {'' }\else $'' $\fi}
\slugcomment{}

\shortauthors{Rhode \& Zepf}
\shorttitle{Globular Cluster Systems of Early-type Galaxies}

\begin{document}

\title{The Globular Cluster Systems of the Early-type Galaxies
NGC~3379, NGC~4406, and NGC~4594\\ and Implications for Galaxy Formation}

\author{Katherine L. Rhode\altaffilmark{1,2}}
\affil{Department of Astronomy, Yale University, New Haven, CT 06520}
\email{rhode@astro.yale.edu}

\author{Stephen E. Zepf\altaffilmark{1}}
\affil{Department of Physics \& Astronomy, Michigan State University, East Lansing, MI 48824}
\email{zepf@pa.msu.edu}

\altaffiltext{1}{Visiting Astronomer, Kitt Peak National Observatory,
National Optical Astronomy Observatories, which is operated by the
Association of Universities for Research in Astronomy (AURA), Inc.,
under cooperative agreement with the National Science Foundation.}
\altaffiltext{2}{NASA Graduate Student Researchers Program Fellow}

\begin{abstract}
We have investigated the global properties of the globular cluster
(GC) systems of three early-type galaxies: the Virgo cluster
elliptical NGC~4406, the field elliptical NGC~3379, and the field S0
galaxy NGC~4594.  These galaxies were observed as part of a wide-field
CCD survey of the GC populations of a large sample of normal galaxies
beyond the Local Group.  Images obtained with the Mosaic detector on
the Kitt Peak 4-m telescope provide radial coverage to at least
24$\arcm$, or $\sim$70$-$100~kpc.  We use $BVR$ photometry and image
classification to select GC candidates and thereby reduce
contamination from non-GCs, and HST WFPC2 data to help quantify the
contamination that remains.  The GC systems of all three galaxies have
color distributions with at least two peaks and show modest negative
color gradients.  The proportions of blue GCs range from 60$-$70\% of
the total populations.  The GC specific frequency ($S_N$) of NGC~4406
is 3.5$\pm$0.5, $\sim$20\% lower than past estimates and nearly
identical to $S_N$ for the other Virgo cluster elliptical included in
our survey, NGC~4472.  $S_N$ for NGC~3379 and NGC~4594 are 1.2$\pm$0.3
and 2.1$\pm$0.3, respectively; these are similar to past values but
the errors have been reduced by a factor of 2$-$3.  We compare our
results for the early-type sample (including NGC~4472) to models for
the formation of massive galaxies and their GC systems.  Of the
scenarios we consider, a hierarchical merging picture --- in which
metal-poor GCs form at high redshift in protogalactic building blocks
and metal-rich GC populations are built up over time during subsequent
gas-rich mergers --- appears most consistent with the data.
\end{abstract}

\keywords{galaxies: individual (NGC~3379, NGC~4406, NGC~4594);
galaxies: star clusters; galaxies: elliptical and lenticular, cD;
galaxies: formation}

\section{Introduction}

Globular clusters (GCs) are an important constituent of galaxies of
all morphological types.  Early-type galaxies such as giant
ellipticals and cD galaxies have long been known to have very populous
GC systems \citep{baum55,sandage61}, often consisting of thousands of
clusters \citep{harris91,az98}.  Much of the work on
extragalactic GC systems has focused on understanding how these
enormous populations of GCs and their host galaxies formed.

Various models have been proposed that either predict or attempt to
explain how it is that ellipticals have such populous GC systems and
also that in many cases broadband colors indicate the presence of two
or more distinct populations of GCs (e.g., Zepf \& Ashman 1993,
Geisler, Lee, \& Kim 1996, Gebhardt \& Kissler-Patig 1999, Larsen et
al.\ 2001, Kundu \& Whitmore 2001).  For example, Ashman \& Zepf
(1992; AZ92) suggested that ellipticals are formed when two more more
disk galaxies merge; GCs in the progenitor spirals make up the
metal-poor population of the elliptical, and its metal-rich GCs are
formed during the merger itself.  Observations of bimodal color
distributions in GCs (e.g., Zepf \& Ashman 1993) and massive star
clusters forming in mergers (e.g., Whitmore et al.\ 1993; Whitmore \&
Schweizer 1995) agreed with earlier predictions of AZ92, lending
support to the model.  Subsequently, other models were proposed to
also account for these observations.  Forbes, Brodie, \& Grillmair
(1997a; FBG97) put forth the idea that metal-poor GCs in ellipticals
formed during the collapse of a protogalactic gas cloud, and
metal-rich GCs originated in a subsequent phase of star formation.
C\^ot\'e, Marzke, \& West (1998; CMW98) asserted instead that it is
the metal-rich GCs that are formed during the initial collapse, and
metal-poor ones are accreted into the elliptical's halo along with the
dwarf galaxies that host them.  More recently, the idea that galaxies
and their globular cluster populations form in galaxy mergers has been
put into a cosmological context by \citet{beasley02} and
\citet{santos03}.  These authors have suggested that metal-poor GCs
are formed at high redshift in protogalactic building blocks; in this
type of scenario, massive ellipticals and their metal-rich GC
subpopulations are built up over time via hierarchical merging.

We initiated a wide-field CCD survey of the GC systems of a sample of
massive galaxies with the goal of improving the observational data so
that it could begin to distinguish between the various models.  In
such a photometric survey, the signature of a GC system around a
galaxy is an overdensity of compact objects with the colors and
magnitudes expected for GCs at that distance.  Past surveys were
carried out before the advent of wide-field, mosaiced CCD imagers, and
typically selected GCs in one or two filters.  As a result, the GC
candidate samples often were significantly contaminated by foreground
stars and background galaxies and many times only a small portion of
the GC system was observed (sometimes less than 10\%; see Appendix in
Ashman \& Zepf 1998).  Large extrapolations were therefore necessary
in order to derive global properties of the GC systems, such as
overall mean colors, total numbers, and specific frequencies, making
these values subject to considerable uncertainties.  Well-constrained
global values are what is needed to test the model predictions as well
as to improve our overall understanding of GC systems and their
relationship to galaxy formation.

Accordingly, we have observed the GC systems of a large sample of both
early-type and spiral galaxies.  The survey employs wide-field CCD
imaging to observe the full extent of the galaxy GC systems, good
resolution to help eliminate background galaxies from the GC candidate
lists, and three-color photometry to isolate {\it bona fide} GCs from
contaminating stars and galaxies. In addition, archival Hubble Space
Telescope (HST) images are available for most of the targets; these
data are used to help quantify the level of contamination in the GC
lists.

In Paper~I of this series \citep{rhode01}, we published our results
for the giant elliptical NGC~4472, the brightest galaxy in Virgo.  We
used NGC~4472 as a test case, in order to refine our techniques for
finding GCs and removing and quantifying contamination.  Paper~II
\citep{rhode03} describes the results for the first spiral to be
fully analyzed, NGC~7814.  Here we present results for the rest of the
early-type galaxy sample: NGC~4406, another giant elliptical in Virgo;
NGC~3379, a less-luminous elliptical in the Leo-I group; and
NGC~4594, a field galaxy that is intermediate between ellipticals and
early-type spirals. Table~\ref{table:Es properties} summarizes the basic
properties of the galaxies in the early-type sample, including
NGC~4472.  We chose galaxies located in different environments and
with a range of luminosities in order to investigate how the
properties of their GC systems depend on these factors.

The paper is organized as follows: Section~\ref{section:Es obs and redux}
summarizes the observations and initial reductions of the data.
Section~\ref{section:Es analysis} describes our methods for detecting and
selecting GC candidates around the target galaxies, as well as our
techniques for quantifying how much of the GC system has been
observed.  Results, including the spatial and color distributions of
the GC candidates and the total number and specific frequency of GCs
for each galaxy, are presented in Section~\ref{section:Es results}.  The
last two sections consist of a discussion of the results and their
implications for galaxy formation models, followed by a summary of our
conclusions.

\section{Observations and Initial Reductions}
\label{section:Es obs and redux}

The early-type galaxy sample was imaged in 1999 March with the Mosaic
Imager on the 4-m Mayall Telescope at Kitt Peak National Observatory
(KPNO).  The Mosaic is made up of eight 2048 x 4096 CCDs separated by
small gaps $\sim$50 pixels wide.  It provides a 36$\arcm$x36$\arcm$
field-of-view when mounted on the Mayall telescope and each pixel
subtends 0.26$\arcs$ of sky.  Five images were taken through each of
three broadband filters ($BVR$) and the telescope was dithered between
exposures to facilitate cosmic ray removal and provide sky coverage in
the gaps between the CCDs. Total integration times were 5400s in $V$
and ranged from 3300s to 3900s in $B$ and 2100s to 2400s in $R$.  All
the images except the $B$ exposures of NGC~3379 were taken under
clear, but not photometric, conditions.  The $B$ observations of
NGC~3379 were taken on a subsequent night under poorer sky conditions,
resulting in a combined $B$ image that is relatively shallow.

Initial data reduction steps were executed using the
IRAF\footnote{IRAF is distributed by the National Optical Astronomy
Observatories, which are operated by AURA, under cooperative agreement
with the National Science Foundation.} package MSCRED.  We executed
the same steps described in Paper~I to process the images (i.e., to
perform overscan and bias subtraction and flat-field division) and
construct a single, stacked image in each of the three filters for
each galaxy.  The resolution (point-spread function FWHM) of the final
stacked images ranges from 1.0$\arcs$ to 1.2$\arcs$ for NGC~4594 and
NGC~4406, and from 1.3$\arcs$ to 1.4$\arcs$ for NGC~3379.

Because the Mosaic data were taken during nonphotometric conditions,
additional calibration data were obtained at KPNO with the WIYN 3.5-m
telescope\footnote{The WIYN Observatory is a joint facility of the
University of Wisconsin, Indiana University, Yale University, and the
National Optical Astronomy Observatories.} and Minimosaic camera on
photometric nights in 2001 April (for NGC~4406) and 2002 March (for
NGC~3379 and NGC~4594).  Single short (400$-$600s) $BVR$ exposures
were taken of each galaxy along with several sets of standard star
frames \citep{land92}.  The Landolt standards were first used to
calibrate the WIYN data. Then $\sim$20 bright but unsaturated stars
were found in common between the WIYN galaxy images and the stacked
Mosaic images.  Magnitudes and colors of these stars were measured in
both sets of images and photometric coefficients
were derived to post-calibrate the Mosaic data to the WIYN data.

\section{Data Analysis}
\label{section:Es analysis}

Below are outlined the steps used to analyze the GC systems of the
three galaxies discussed in this paper; see Paper~I for a more
detailed description of our methods.

\subsection{Source Detection and Matching}

To locate GC candidates in the stacked Mosaic images of each galaxy,
we smoothed the images with a ring median filter of diameter 6 times
the mean FWHM of point sources in the image.  Subtracting the smoothed
images from the originals removed the galaxy light while preserving
the flux from the discrete sources.  A constant background level was
added to the galaxy-subtracted images and DAOFIND was used to detect
all sources above a specified threshold (typically, between three and
five times the noise in the local background).  A final list of
objects appearing in all three filters was produced; the total number
of detected sources was 3224 in NGC~3379, 12502 in NGC~4406, and 8053
in NGC~4594.

\subsection{Extended Source Cut}
\label{section:Es ext src cut}

\begin{figure}
\plotone{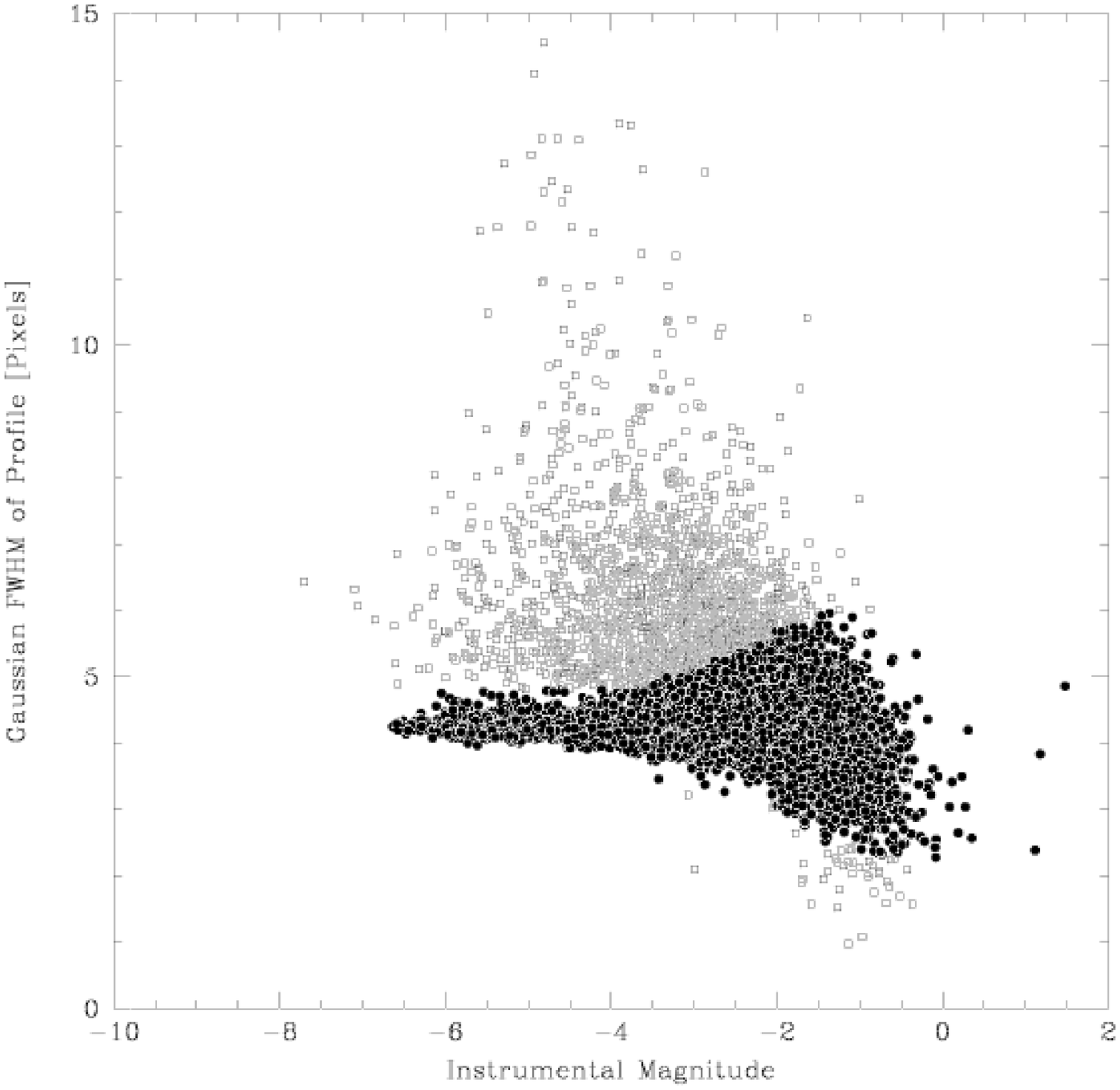}
\caption{Gaussian FWHM of the radial profile versus instrumental
magnitude for 8053 objects in the stacked Mosaic $V$ image of
NGC~4594.  Open circles are objects that are deemed extended and that
were subsequently eliminated from the list of GC candidates.  Filled
circles mark the objects that were accepted as possible point sources.
\protect\label{fig:fwhm mag}}
\end{figure}

To eliminate contaminating background galaxies from the source lists,
we measured the FWHM and instrumental magnitude of each source.
Objects with FWHM values larger or smaller than expected for a point
source of a given magnitude were rejected.  As an example, FWHM versus
magnitude for 8053 sources in the Mosaic $V$-band image of NGC~4594 is
plotted in Figure~\ref{fig:fwhm mag}.  Open squares are sources deemed
extended and filled circles are those that passed the extended source
cut.  As the figure demonstrates, the range of acceptable FWHM values
increases with increasing instrumental magnitude.  For NGC~4406 and
NGC~4594, we used measurements in all three filters to decide whether
a source was extended.  The $B$ image of NGC~3379 had poorer
resolution so only the $V$ and $R$ images were used.
Typically, 30$-$50\% of the objects in the original source lists were
removed in this step; the number of accepted sources was 1728 in
NGC~3379, 6604 in NGC~4406, and 5708 in NGC~4594.

\subsection{Photometry}

Photometry with an aperture of radius equal to the average FWHM of the
images was performed for objects that passed the extended source cut.
Calibrated $B$, $V$, and $R$ magnitudes were then computed using the
appropriate aperture correction and photometric calibration
coefficients.  The aperture corrections, which were derived in the
same way as for Paper~I, are listed in Table~\ref{table:Es aper corr};
the values given are the difference between the total magnitude and
that within the small aperture.  The final magnitudes were corrected
for Galactic extinction using reddening data from Schlegel, Finkbeiner
\& Davis (1998); the extinction corrections are given in
Table~\ref{table:Es ext corr}.

\subsection{Color Selection}
\label{section:Es color cut}

Observations in three filters can help separate GCs from contaminating
background galaxies because in the $BVR$ color-color plane, GCs are
well-separated from early-type spirals and ellipticals and
high-redshift galaxies.  They overlap low-redshift, late-type spirals
and moderate-redshift irregulars, but many of the low-$z$ objects are
eliminated in the extended source cut.  Thus combining three-filter
photometry with image analysis substantially reduces the level of
contamination in our GC candidate lists.

The color selection was executed as follows: objects with absolute
magnitude brighter than $M_V$ $=$ $-$11 (assuming the galaxy distances
listed in Table~\ref{table:Es properties}) were eliminated.  A
photometric error threshold of 0.125 to 0.15 magnitudes in one or more
filters (depending on the quality of the images) was also imposed.
Finally, taking into account their photometric errors, objects with
$B-V$ between 0.56 and 0.99 (corresponding to [Fe/H] of $-$2.5 to 0.0
for Galactic GCs) and $V-R$ values that put them within a specified
distance from the relation between $B-V$ and $V-R$ for Galactic
GCs were retained.
The distance from the $B-V$ versus $V-R$ line that was deemed
acceptable varied. For NGC~4406 and NGC~4594, the data were of good
quality and a 2$-\sigma$ distance above and below the line (where
$\sigma$ is the scatter in the relation for Milky Way GCs) was
sufficient.
The $B$ image of the NGC~3379 field was very shallow and the detected
sources were preferentially blue; for this reason we accepted objects
that were within 1-$\sigma$ above the line (redder than Milky Way GCs)
and 3-$\sigma$ below it (bluer than Milky Way GCs).

Two of the galaxies in the early-type sample have companion galaxies
that appear in the Mosaic images.  NGC~3384 is an SB0 galaxy in the
Leo group and is located only 7$\arcm$ (22~kpc in projection) away
from NGC~3379.  NGC~4374, a Virgo cluster E1 with $M_V$ $=$ $-$22.05,
is situated $\sim$17$\arcm$ (75~kpc in projection) west of NGC~4406.
In both cases these galaxies' GC systems were readily apparent in the
images as surrounding overdensities of point sources with GC-like
colors.  In order to eliminate these objects from the final lists of
GC candidates, we marked a region around the companion galaxies and
removed any source located within that region.  These areas were also
excluded from the spatial coverage calculation described in
Section~\ref{section:Es radial profiles}.

\begin{figure}
\plotone{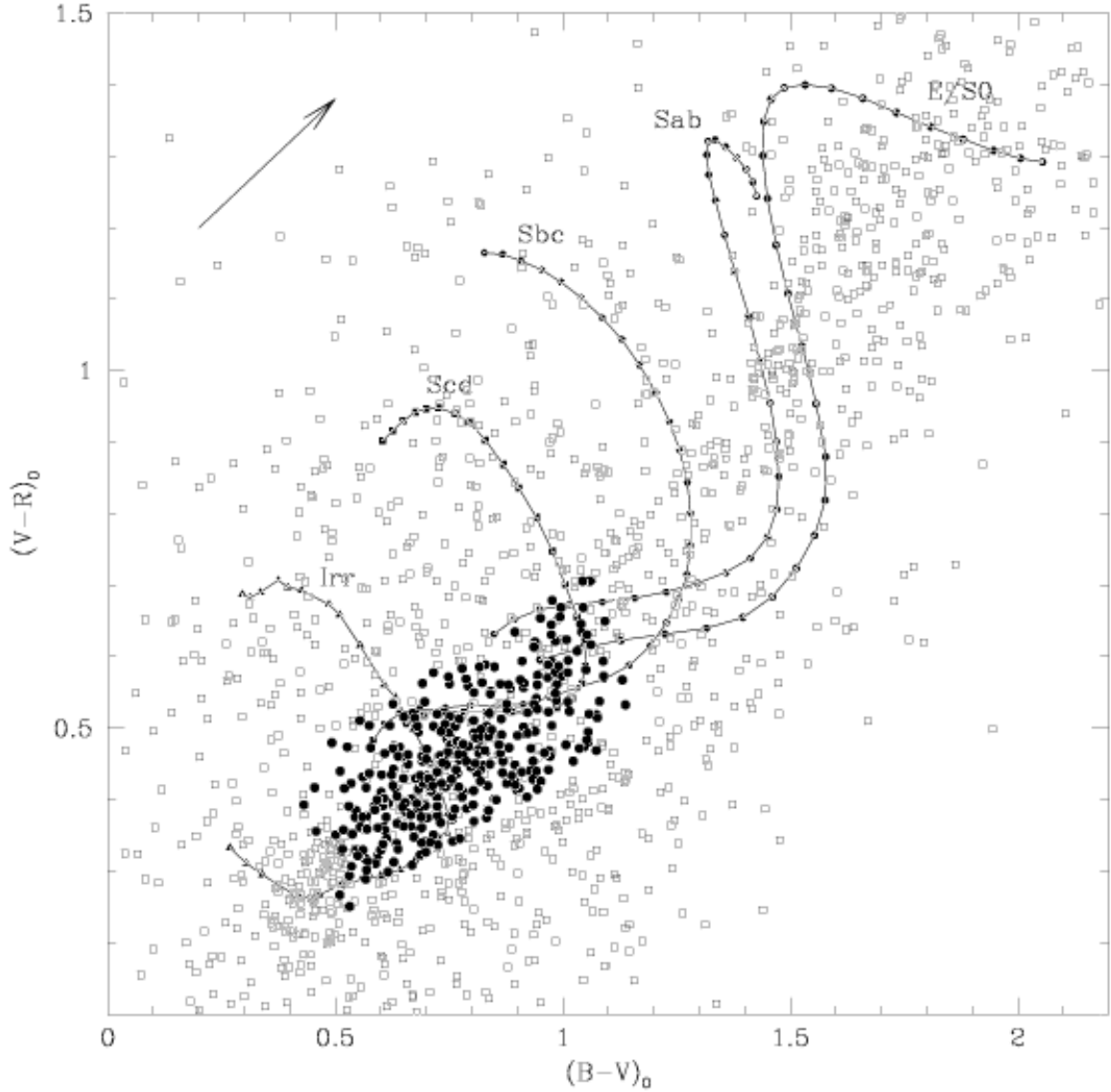}
\caption{Color selection of GC candidates in NGC~3379. Open
squares are 1728 unresolved objects in the Mosaic images and filled circles
are the final set of 321 GC candidates around NGC~3379.  For
reference, the locations in the $BVR$ plane of galaxies of various
types are shown as tracks the galaxies would follow with increasing
redshift.  See Paper~I for details on how the tracks were
produced.  A reddening vector of length $A_V$ $=$ 1~mag appears in the
upper left corner.
\protect\label{fig:color cut n3379}}
\end{figure}

\begin{figure}
\plotone{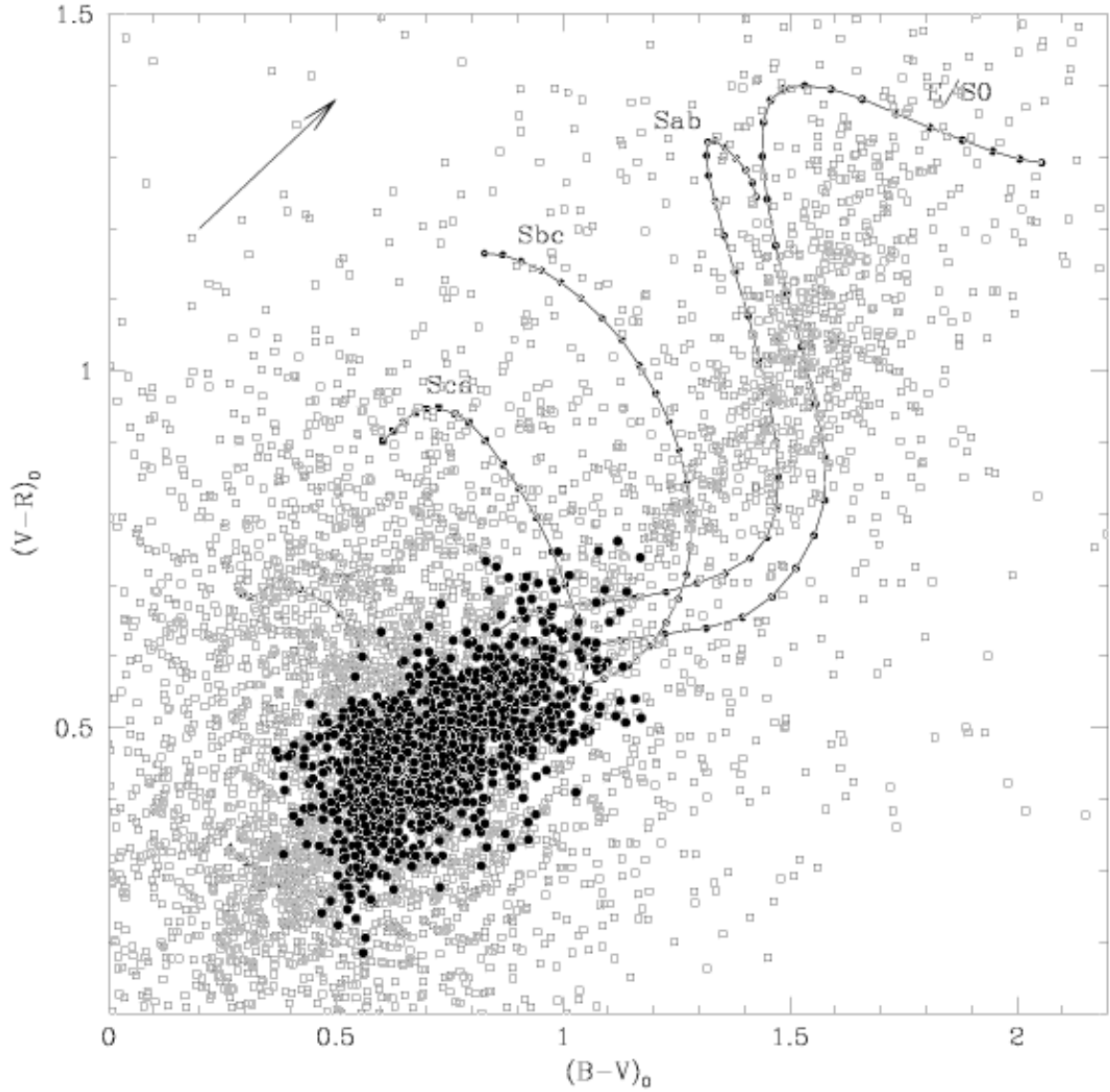}
\caption{Color selection of GC candidates in NGC~4406. Open
squares are 6604 unresolved objects in the Mosaic images and filled
circles are the final set of 1400 GC candidates.
\protect\label{fig:color cut n4406}}
\end{figure}

\begin{figure}
\plotone{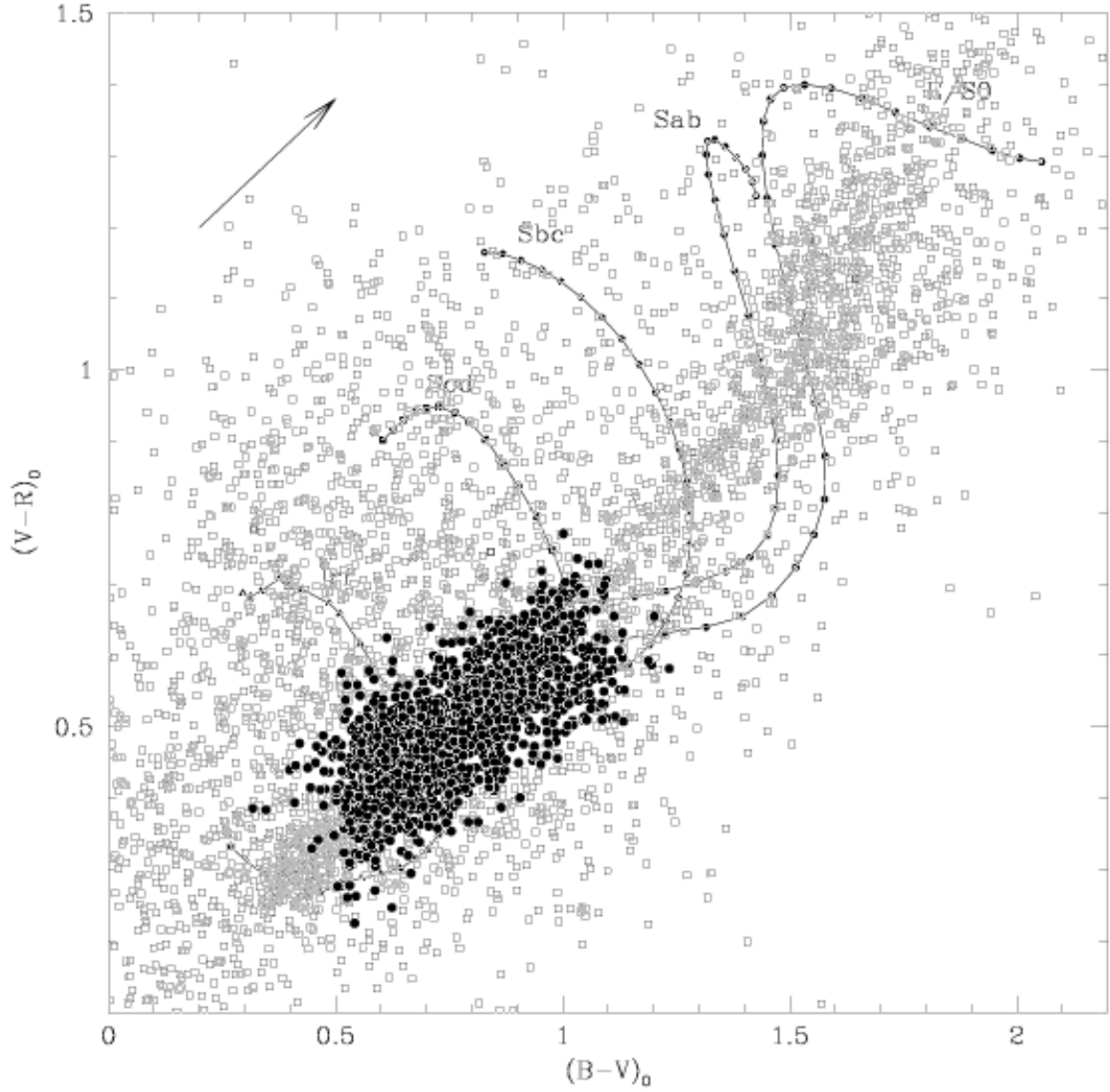}
\caption{Color selection of GC candidates in NGC~4594. Open
squares are 5708 objects that passed the extended source cut and
filled circles are the final set of 1748 GC candidates.
\protect\label{fig:color cut n4594}}
\end{figure}

The final lists of GC candidates around NGC~3379, NGC~4406, and
NGC~4594 contained 321, 1400, and 1748 objects, respectively.
Figures~\ref{fig:color cut n3379} through \ref{fig:color cut n4594}
show the results of the color selection.  The locations in the $BVR$
color-color plane of sources that passed the extended source cut are
shown as open squares, and the GC candidates are plotted as filled
circles.

\subsection{Completeness Testing}
\label{section:Es completeness}

To quantify the detection limits of the Mosaic images, a series of
completeness tests was executed for each galaxy.  First, 800
artificial point sources with magnitudes within 0.2-mag of a specified
mean value were added to the $B$, $V$, and $R$ images. Then the same
detection steps executed on the original images were performed and the
fraction of artificial stars recovered in the detection process was
recorded.  The process was repeated over a range of 3$-$5 magnitudes
for each filter.  The 50\% completeness limits are listed in
Table~\ref{table:Es completeness}.  As mentioned earlier, the $B$ image
of NGC~3379 was taken under poor sky conditions and is relatively
shallow.  Consequently the $B$ detection limit dominates the
completeness corrections and the fractional GCLF coverage calculated
in Section~\ref{section:Es gclf fitting} is smaller than intended,
despite NGC~3379's relative proximity.
  
\subsection{Quantifying and Correcting for Contamination}
\label{section:Es contamination}

\subsubsection{Galaxies}
\label{section:Es galaxy contamination}

Compact galaxies that appear as point sources in ground-based data are
often resolved in images taken with the Wide-Field and Planetary
Camera~2 (WFPC2) on HST.  Therefore by locating a subset of our GC
candidates in archival WFPC2 images and determining how many are
galaxies, we can estimate the level of galaxy contamination that
exists in the data.

WFPC2 observations in broadband filters of fields within 15$\arcm$ of
the centers of NGC~3379, NGC~4406, and NGC~4594 were retrieved from
the HST archive\footnote{Based on observations made with the NASA/ESA
{\it Hubble Space Telescope}, obtained from the data archive at the
Space Telescope Science Institute.  STScI is operated by AURA, under
NASA contract NAS 5-26555.}.  ``On-the-fly'' calibration was applied
to the images.  Table~\ref{table:Es hst data} summarizes the data sets
we analyzed.  Column (1) gives HST proposal ID; column (2) is the
target name (names like ``Parallel Field'' and ``Any'' are
observations done by WFPC2 while another instrument was being used for
the primary science); column (3) is the Principal Investigator; column
(4) is the angular separation of the pointing from the center of the
target galaxy; and column (5) is the filter.  When possible, combined
images were created with the STSDAS task CRREJ.  WCSTRAN was used to
locate the Mosaic GC candidates in the WFPC2 frames.  Following the
method of \citet{kundu99}, we measured the flux of the candidates in
two apertures to decide which were galaxies.  Kundu et al.\ found that
at the distance of Virgo, objects in a WF chip with counts$_{3
pix}$/counts$_{0.5 pix}$ $>$ 8 are extended.  (None of our GC
candidates appeared in the PC chip.)  This criterion was used for the
NGC~4406 data.  For NGC~3379 and NGC~4594, we determined that sources
with count ratios $>$ 10 and 12, respectively, were galaxies.  We also
visually inspected the GC candidates to confirm the results from
photometry.

The final statistics are as follows: one of 23 GC candidates appearing
in the WFPC2 images of NGC~3379 is a galaxy; two of 57 GC candidates
in NGC~4406 are galaxies; and five of 43 GC candidates in NGC~4594 are
galaxies.  Combining the number of galaxies with the usable area
covered by the WFPC2 frames yields the estimated number density of
galaxies in our samples.  The number densities for NGC~3379, NGC~4406
and NGC~4594 are, respectively, 0.08 per square arc minute, 0.16 per
square arc minute, and 0.26 per square arc minute.

\subsubsection{Stars}
\label{section:Es stellar contamination}

We ran the Galactic structure model code from \citet{mendez96} and
\citet{mendez00} to estimate how many foreground stars are included in
the GC candidate lists.  Given values for parameters such as the solar
distance and the proportion of stars in the various components of the
Galaxy,
the model calculates the surface density of Galactic stars in a given
magnitude and color range in a specific direction on the sky.  The
model predicts that in the direction of NGC~3379, the number density
of stars with $V$ magnitudes and $B-V$ colors in the same range as the
GC candidates is 0.11 per square arc minute.  The answer was the same
for NGC~4406.  For NGC~4594, the stellar surface density was 0.28 per
square arc minute.  The results were relatively insensitive to changes
in our choice of the composition of the Milky Way and the location of
the Sun within the disk.

\subsubsection{Radially-Dependent Contamination Correction}
\label{section:Es radial contam}

The estimated fraction of contaminating objects at each radius around
the target galaxies was calculated for use in subsequent steps.  The
GC candidates in each galaxy were assigned to annuli of width
1$\arcm$, with the first annulus centered at $r$ $\sim$ 1$\arcm$ and
the last at $r$ $\sim$ 24$\arcm$ for NGC~3379 and NGC~4406, and at $r$
$\sim$ 25$\arcm$ for NGC~4594.  The number of contaminating objects in
each annulus was calculated by multiplying the number density of stars
plus galaxies by the effective area (the region in which GCs could
actually be detected; see Section~\ref{section:Es radial profiles}) of
the annulus. Dividing the number of contaminating objects by the
number of GC candidates in each annulus yielded the contamination
fraction as a function of radius.
NGC~4594 and NGC~4406 have large numbers of GC candidates so the
contamination fraction is modest at radii inside 7$-$10$\arcm$ (e.g.,
in NGC~4406 it is 20\% by 10$\arcm$, or $\sim$45~kpc).  The GC system
of NGC~3379 is more sparsely populated, so the estimated contamination
fraction climbs quickly in the inner few arc minutes, already reaching
$\sim$40\% by 4$\arcm$ ($\sim$12~kpc).

\subsection{Determining the GCLF Coverage}
\label{section:Es gclf fitting}

An observed GC luminosity function (GCLF) for each galaxy was
constructed by assigning the $V$ magnitudes of the GC candidates to
bins of width 0.3 mag.  The radially-dependent contamination
correction 
was applied during the binning process.
The luminosity function was corrected for completeness by computing
the total completeness of each $V$ bin (which involves convolving the
completenesses of all three filters, as described in Paper~I) and
dividing the number of GCs in the bin by this value.

We assumed that the intrinsic GCLFs of the three early-type galaxies
had peaks at $M_V$ $=$ $-$7.4, which is consistent with results from
HST studies of ellipticals \citep{kw01}.  Combining this with the
distance moduli listed in Table~\ref{table:Es properties} gives $V$ $=$
22.7, 23.7 and 22.6 for the peak apparent magnitude of the GCLF for
NGC~3379, NGC~4406 and NGC~4594, respectively.  We constructed
Gaussian distributions with these peak magnitudes and dispersions of
1.2, 1.3 and 1.4 mag and fitted them to the observed luminosity
functions, varying the normalization of the Gaussian. Bins with less
than 45\% completeness were excluded from the fits.  The fraction of
the best-fit theoretical GCLF covered by our data was calculated for
each case.

\begin{figure}
\plotone{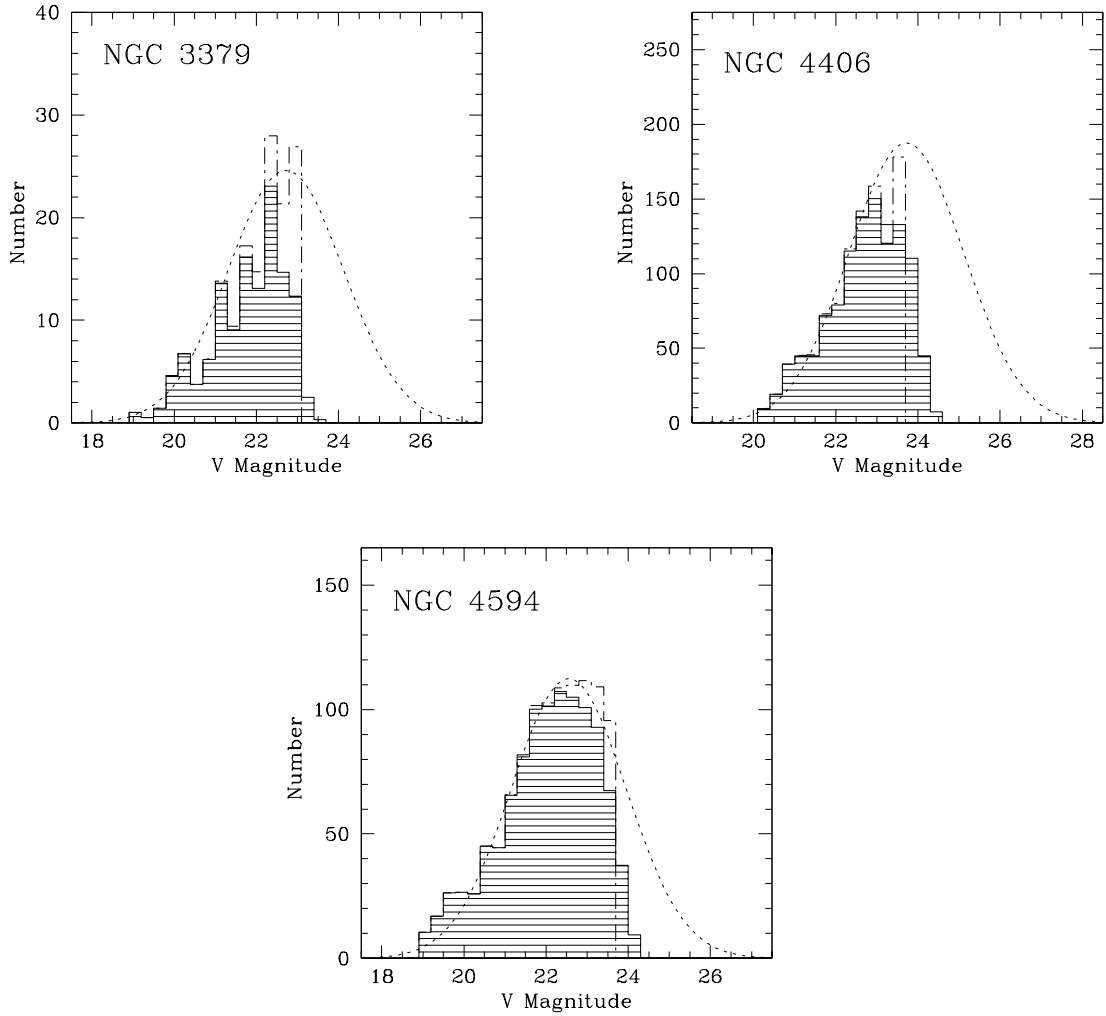}
\caption{GCLF fitting for the three galaxies.  The shaded histogram is
the observed GC luminosity function.  The dashed line shows the
completeness-corrected data used in the fit (i.e., bins with
completeness $>$45\%).  The dotted line is the best-fit Gaussian
distribution.  \protect\label{fig:Es gclfs}}
\end{figure}

The quality and depth of the observed LF data vary for the three
galaxies.  For NGC~3379, the data do not reach the turnover of the
assumed GCLF and several of the histogram bins are sparsely populated.
Different dispersions produced similar-quality fits.  The mean
fractional coverage derived from all three fits is 0.46$\pm$0.01.  The
NGC~4406 data also do not reach the peak of the GCLF, but the bins
leading up to it are well-populated.  The Gaussian with 1.4-mag
dispersion clearly yielded the best fit; the associated fractional
coverage is 0.51. (Smaller dispersions produced values within 1\% of
this.)  For NGC~4594, the observed data go well past the peak of the
GCLF and are well-fit by a Gaussian with 1.4-mag dispersion and
inconsistent with smaller dispersions.  The fractional coverage is
0.83.  Figure~\ref{fig:Es gclfs} shows the GCLF fits for the three
galaxies. The shaded histogram is the observed data; the dashed
histogram has been corrected for completeness; and the dotted lines
are the fitted Gaussian distributions with the lowest $\chi^2$ values.

We investigated whether changes in the bin size of the LF data
affected the fractional coverage results for each galaxy.  For
NGC~3379 and NGC~4406, changing the bin size changed the fractional
coverage by as much as 2.5\%.  For NGC~4594, the observed data covered
so much of the GCLF that the change was only a few tenths of a
percent.  These uncertainties are taken into account in the error
calculation in Section~\ref{section:Es spec freq}.

\section{Results}
\label{section:Es results}
\subsection{Radial Distributions}
\label{section:Es radial profiles}

\begin{figure}
\plotone{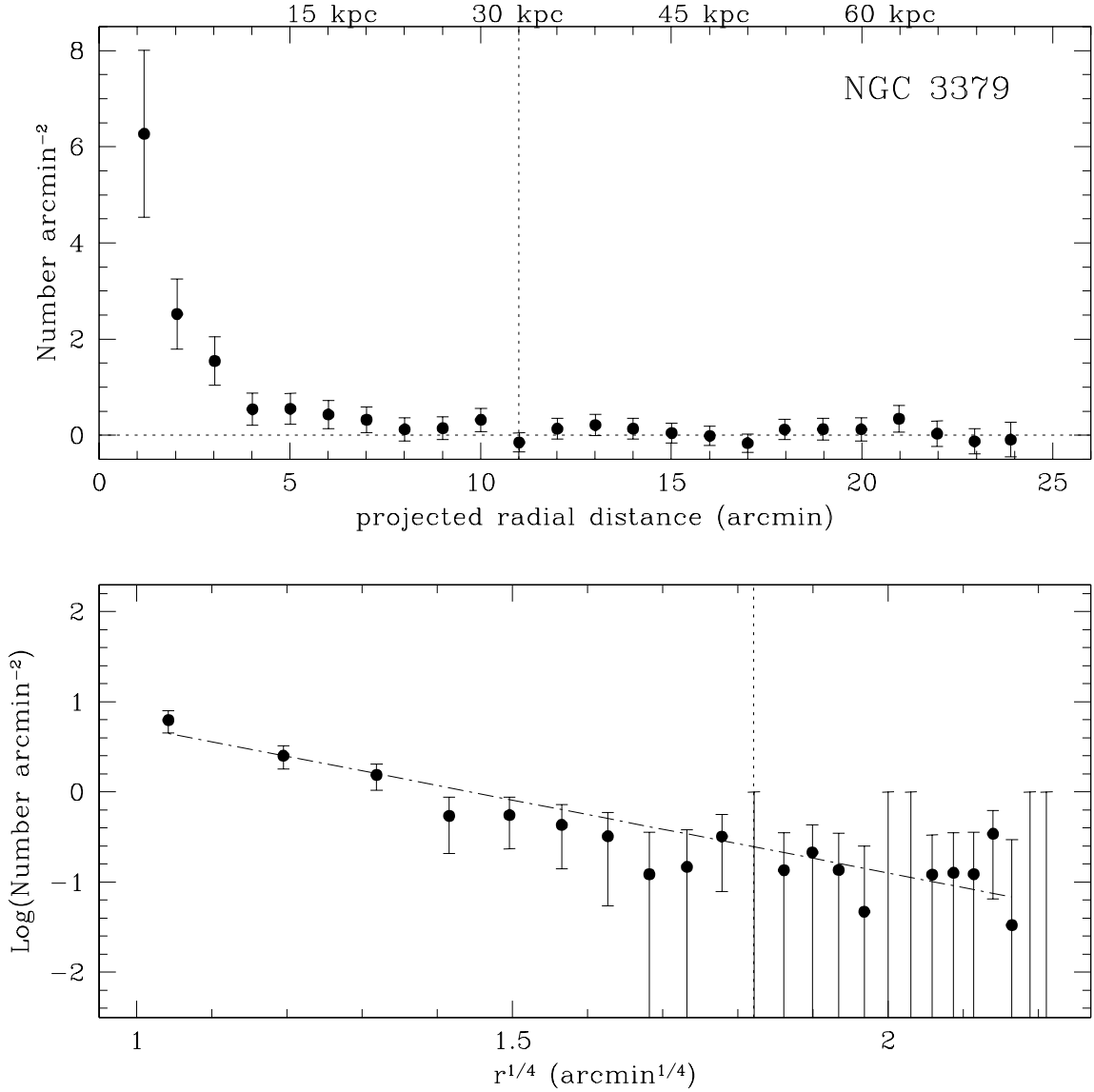}
\caption{Radial distribution of GCs in NGC~3379, plotted as surface
density vs.\ projected radial distance (top) and as the log of the
surface density vs.\ $r^{1/4}$ (bottom).  The dotted line in the top
plot indicates zero surface density and the dashed line in the bottom
plot is the best-fit deVaucouleurs law.  The vertical line in both
plots marks the location where the surface density is consistent with
zero within the errors.  The data have been corrected for
contamination, missing spatial coverage, and magnitude incompleteness.
\protect\label{fig:profile n3379}}
\end{figure}

\begin{figure}
\plotone{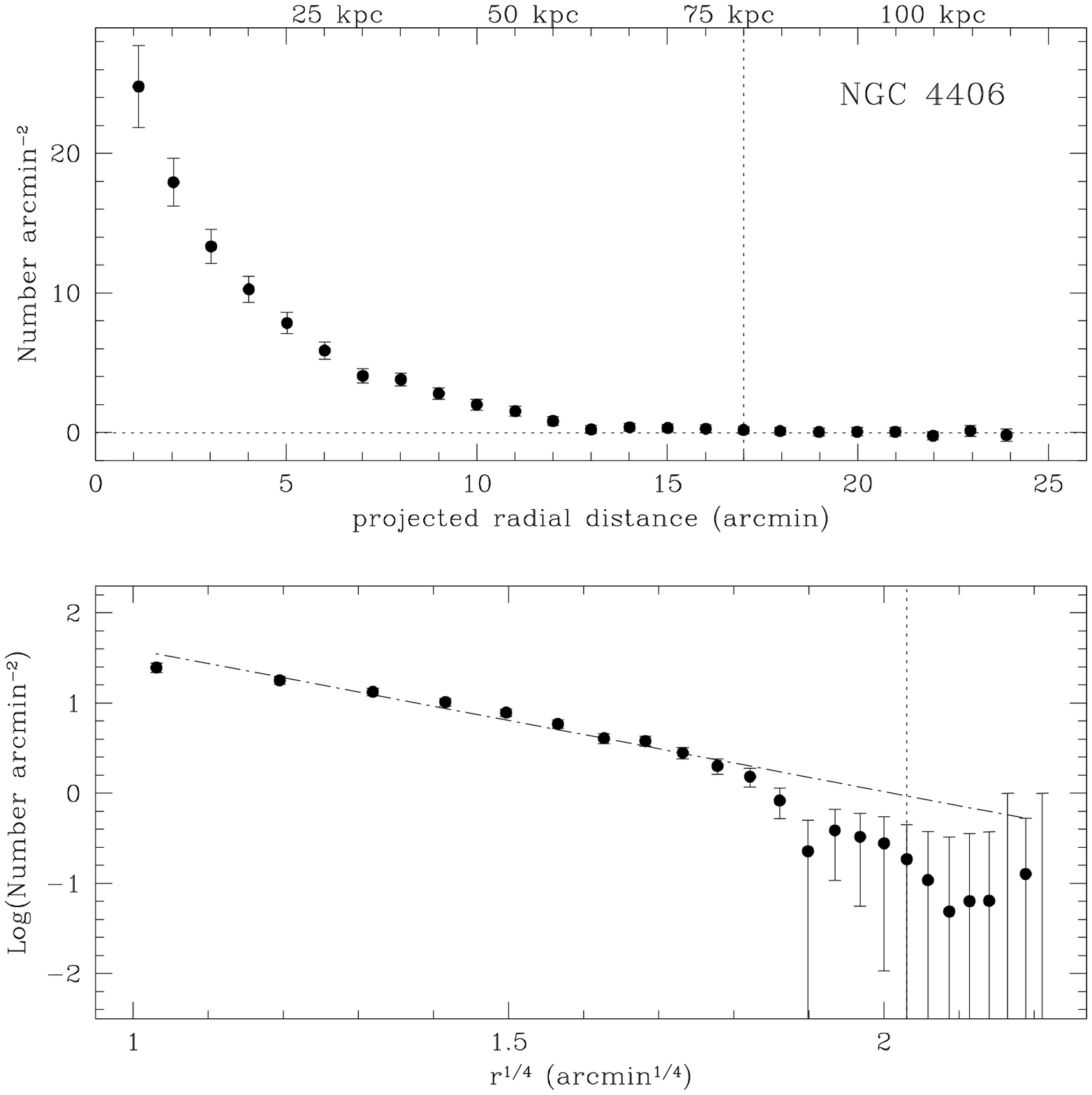}
\caption{Radial distribution of GCs in NGC~4406, plotted in the same
way as in Figure~\ref{fig:profile n3379}.
\protect\label{fig:profile n4406}}
\end{figure}

\begin{figure}
\plotone{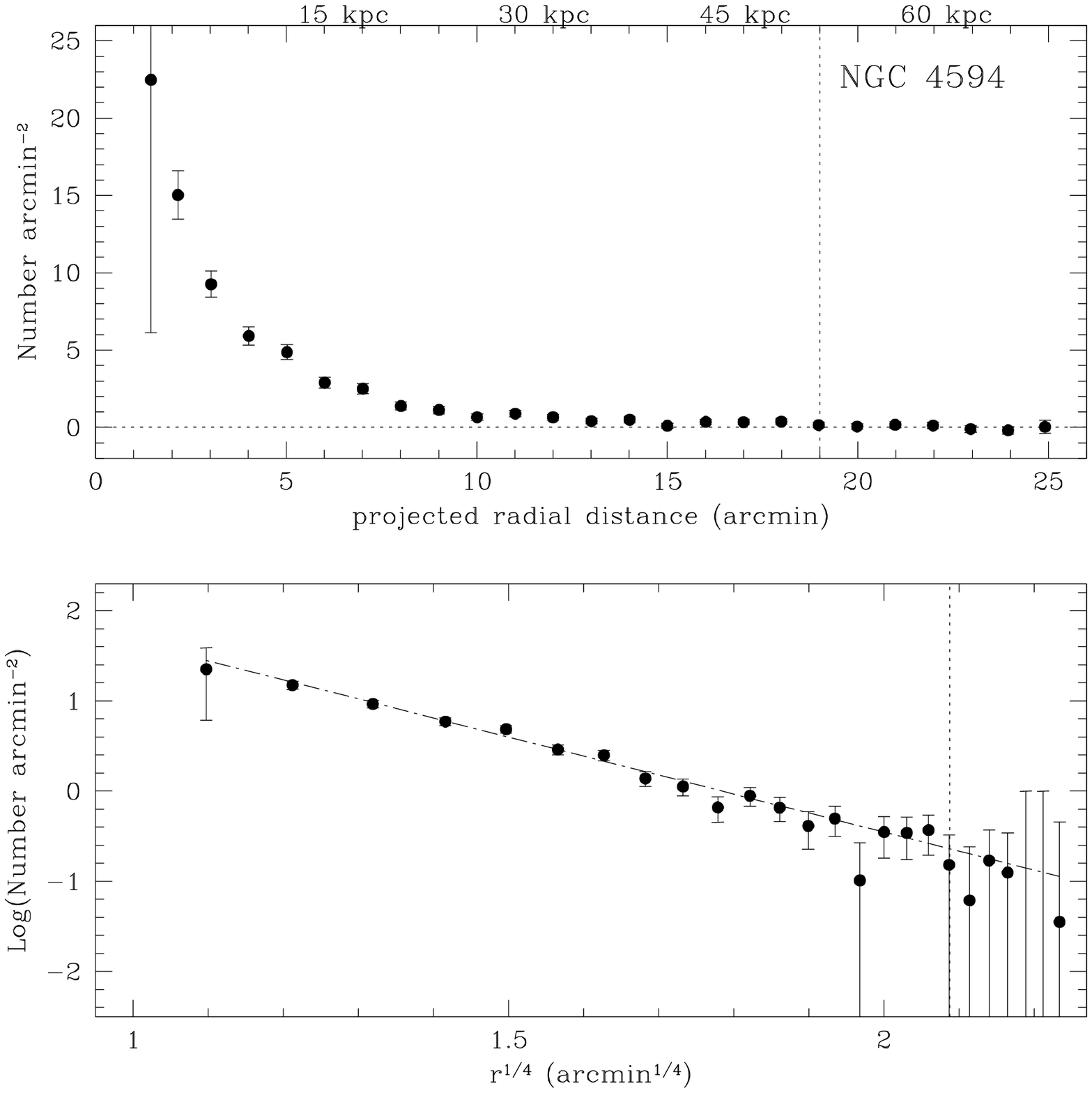}
\caption{Radial distribution of GCs in NGC~4594, plotted in the same
way as in Figure~\ref{fig:profile n3379}.
\protect\label{fig:profile n4594}}
\end{figure}

To investigate the spatial distribution of the GC systems, GC
candidates were assigned to 1$\arcm$-wide annuli based on their
projected radial distances from the galaxy centers.  An effective area
was calculated for each annulus that excluded regions where GCs could
not be detected (i.e., regions around bright stars and companion
galaxies that had been masked out, the central $\sim$0.5$\arcm$ of
each galaxy, and parts of the annulus that extended off the image).
The number of GCs in each annulus was corrected for contamination and
GCLF coverage. The corrected radial distributions are shown in
Figures~\ref{fig:profile n3379} through \ref{fig:profile n4594} and
the numerical values are listed in Table~\ref{table:Es profiles}.  Column
(1) of the table gives the mean radius of the unmasked pixels in each
annulus, column (2) lists the surface density and associated error,
and column (3) gives the fraction of the annulus with spatial
coverage.

Each GC profile is different, but in all three cases the surface
density decreases until it is consistent with zero in several adjacent
bins, well before the end of the data.  This suggests that we have
observed the full radial extent of the galaxies' GC systems.  NGC~3379
has the least populous GC system and the surface density is consistent
with zero within the 1-$\sigma$ error bars by the 11$\arcm$ bin
($\sim$30~kpc) and beyond. (It becomes slightly positive in one outer
bin at 21$\arcm$, but returns to zero or negative values in the next
three annuli.)  In NGC~4406, the surface density is consistent with
zero by 17$\arcm$ ($\sim$75 kpc) outward.  NGC~4594's GC surface
density decreases to very low values by 15$\arcm$ and is zero within
the errors in every bin from 19$\arcm$ ($\sim$50 kpc) to the end of
the data at 25$\arcm$.  The innermost point in the radial profile for
NGC~4594 has comparatively large error bars because nearly the entire
annulus was masked out due to the galaxy's dust lane.  Only two GCs
were detected in the small region left unmasked.  We have included the
point in the figure and table for the sake of completeness, but
removing it does not change the results.

We fitted the GC spatial distributions with deVaucouleurs law profiles
of the form log~$\sigma_{\rm GC}$ $=$ $a0$ $+$ $a1$~$r^{1/4}$ and
power laws of the form log~$\sigma_{\rm GC}$ $=$ $a0$ $+$
$a1$~log~$r$; the results are given in Table~\ref{table:Es coefficients}.
For NGC~4594 and NGC~4406, the deVaucouleurs law provides a better fit
than the power law at large radius, where the data begin to drop
slightly or significantly below the best-fit lines.  This is similar
to NGC~4472, for which a deVaucouleurs law also fits the GC radial
profile better than a power law.  For NGC~3379, both the power law and
deVaucouleurs law provide good fits out to the limit of the data; the
best-fit lines intersect nearly all the points plus error bars. For
consistency, we have shown the deVaucouleurs law fits in the figures
for all three galaxies (including NGC~3379) and we use them to
calculate the total number of GCs for each galaxy in
Section~\ref{section:Es spec freq}.

The behavior of NGC~4406's radial distribution is somewhat different
from that of the other two galaxies.  The data for NGC~3379 and
NGC~4594 scatter around the best-fit deVaucouleurs law profile out to
the point at which the surface density becomes consistent with zero
(marked with a vertical line in the figures), and intersects the error
bars of each point in most of the bins inside that line.  For
NGC~4406, however, the data start to fall below the best-fit line
around 10$-$11$\arcm$, and by 12$\arcm$ ($\sim$50~kpc) onward, they
lie systematically below the line.  This is not necessarily unexpected
behavior (since there is no reason to assume that a deVaucouleurs law
{\it must} provide a good fit over the entire extent of a GC system)
but is worth noting in contrast to the other galaxies.  NGC~4406 is in
a richer environment than NGC~3379 and NGC~4594 and has a massive,
close companion; it is perhaps possible that its GC system is showing
signs of tidal truncation either due to interaction with NGC~4374 or
the tidal field of the Virgo cluster.

Although we had masked out NGC~4374 and its GC system when
constructing the radial profile for NGC~4406, we wanted to confirm
that the observed behavior of the profile was real and not due to the
presence of GCs from NGC~4374 artificially inflating the surface
density in the inner few arc~minutes.  (The masked region begins in
the 10$\arcm$ bin, near where the data start to fall below the
best-fit line.)  To test this we constructed a radial profile for
NGC~4406 using only the side of the galaxy away from NGC~4374, i.e.,
the eastern half of NGC~4406's GC system.  The extent of this profile
is similar to that of the original one; the surface density in the
eastern profile is consistent with zero by 15$\arcm$ outward, whereas
the original profile has a slightly positive surface density
(0.28$\pm$0.27 per square arc~minute) in the 16$\arcm$ bin and then is
consistent with zero at larger radius.  The slope and intercept of the
best-fit deVaucouleurs law are the same for both profiles. Furthermore,
the behavior of the eastern-half profile is similar to that of the
original profile: the data plus error bars once again fall
systematically below the best-fit line from 12$\arcm$ onward.
Therefore it appears that the behavior seen in the original profile is
a real effect and not simply due to inadequate masking of NGC~4374's
GC system.

\subsection{Color Distributions}
\label{section:Es color distributions}

To investigate the colors of the GCs in each galaxy, we selected GC
candidate samples that are at least 90\% complete in all three
filters.  For example, the reddest GC candidate in NGC~3379 has $B-R$
$=$ 1.8 and our detection is 90\% complete at $B$ $=$22.5, so
candidates with $R$ brighter than 20.7 were chosen.  Any GC candidates
found to be galaxies in the HST images were excluded. We also imposed
a radial cut at the location where the GC surface density in the
radial profile became consistent with zero.  This yielded samples of
36, 979, and 1084 objects in NGC~3379, NGC~4406, and NGC~4594,
respectively.

\begin{figure}
\plotone{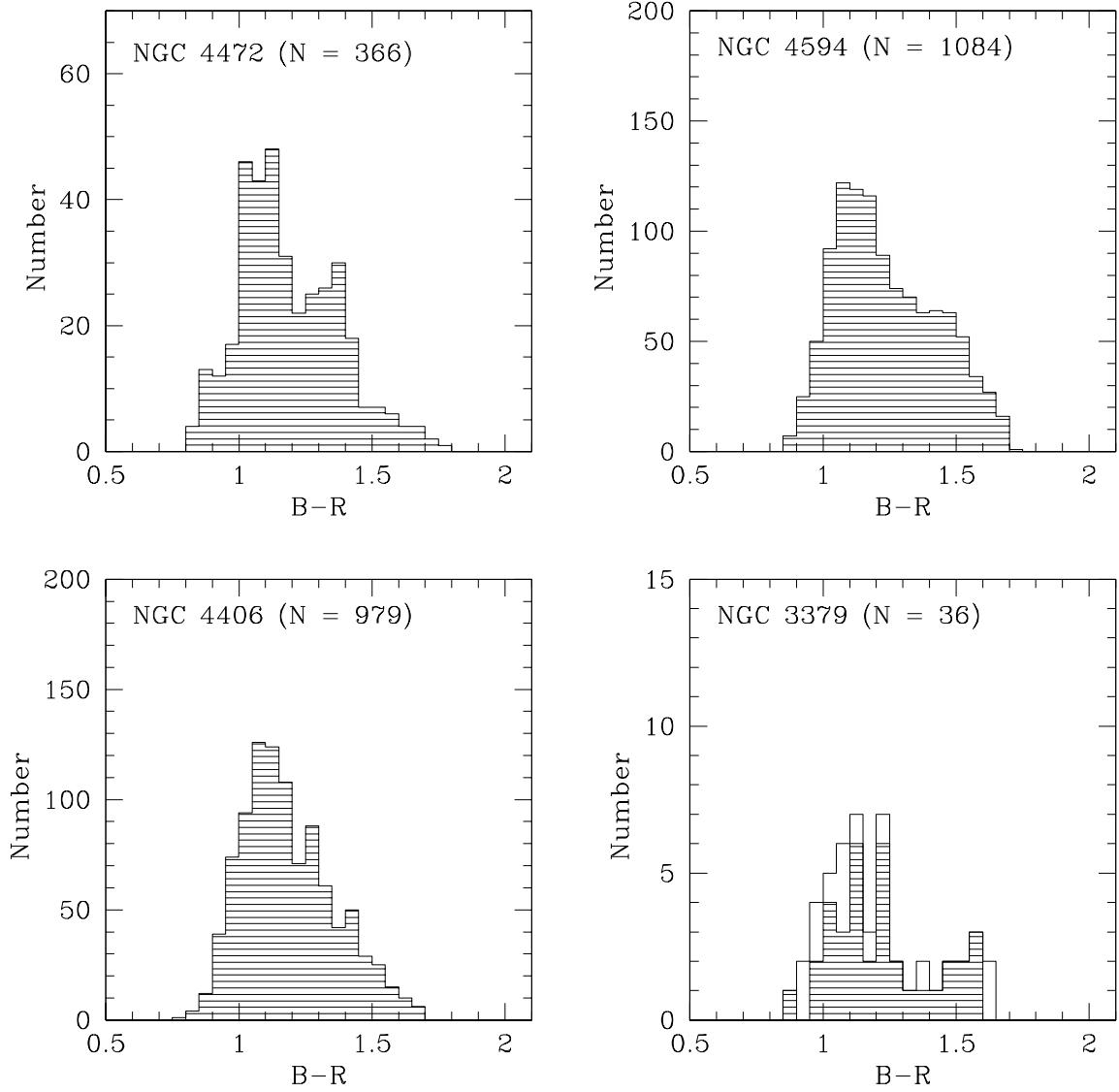}
\caption{$B-R$ distributions for the early-type galaxy sample,
including NGC~4472 from Paper~I.  For NGC~3379, the 36-object sample
used to estimate the blue/red GC proportions is shown as a shaded
histogram and the 50-object sample used as input to KMM is plotted
with a solid line.  
\protect\label{fig:Es color distns}}
\end{figure}

$B-R$ histograms for the early-type galaxy sample are shown in
Figure~\ref{fig:Es color distns}, including that for NGC~4472 from
Paper~I.  NGC~4472 has a bimodal distribution with peaks at $B-R$
$\sim$1.1 and 1.4 and a gap around 1.2; roughly 60\% of the GCs are in
the blue peak and 40\% are red.  The data for NGC~4406 show an obvious
peak at $B-R$ $\sim$ 1.1 and a tail of red GCs, with perhaps hints of
secondary peaks around 1.3 and 1.4. NGC~4594's distribution is wide
and, like NGC~4472's, roughly resembles two overlapping Gaussians.
With only 36 GCs included, NGC~3379's color distribution is too sparse
to show clear peaks, but does not appear consistent with a single
Gaussian.

We ran the KMM algorithm \citep{abz94} on the color distributions,
which tests whether a single or multiple Gaussian functions provide a
better fit to a distribution.  The user specifies the number of
Gaussians to fit as well as whether to use the same or different
dispersions for each one.  The results for each galaxy are as follows:

\noindent{\it NGC 3379.} KMM requires at least 50 objects to produce a
reliable result, so the 36-object sample shown in Figure~\ref{fig:Es
color distns} is too small.  Instead we created a sample with exactly
50 objects by taking the 90\% sample and imposing a slightly relaxed
radial cut --- i.e., including objects with radii inside 13.5$\arcm$,
instead of the 11$\arcm$ criterion used to create the 36-object
sample.  This 50-object sample was used as input to KMM.  The unimodal
hypothesis was rejected for this sample at 99.8\% confidence and peaks
were located at $B-R$ $\sim$1.1 and 1.5.  The separation between the
two components of the fitted distribution was around $B-R$ of 1.3.  If
we split the final 36-object color sample at $B-R$ $=$ 1.3, we find
that 70\% of the GC candidates are bluer than this color and 30\% are
redder.  (Note that the color distributions of both the 36-object and
50-object samples are shown in Figure~\ref{fig:Es color distns}.)

\noindent{\it NGC 4406.}  Fitting with two Gaussians yielded peaks at
$B-R$ $\sim$ 1.1 and 1.4.  The proportion of blue GCs was 56\% if we
chose varying dispersions and 75\% if we required them to be the same.
Since NGC~4406's color distribution may show more than two peaks, we
also fitted it with three Gaussians.
Peaks were located at $B-R$ $\sim$ 1.1, 1.3, and 1.5.  Between 56 and
58 percent of the GCs were assigned to the first (blue) peak and the
remainder to the second two peaks.  In every case, the distribution
was found to be multimodal at 99.99\% confidence.  The separation
between the GC candidates in the bluest peak and those in the redder
peak (or peaks) occurred around $B-R$ of 1.2$-$1.3.

\noindent{\it NGC 4594.} The single-Gaussian hypothesis was rejected
with 99.99\% confidence and a fit with two Gaussians produced peaks at
$B-R$ $\sim$ 1.1 and 1.5.  For Gaussians with identical dispersions,
66\% of the GCs were blue; if the dispersions varied, the proportion
was 59\%.  The separation between the blue and red populations occurs
at $B-R$~$\sim$~1.3.

NGC~4472, NGC~4406, and NGC~4594 have similar proportions of blue GCs
($\sim$60\%) whereas NGC~3379 appears to have a slightly larger blue
population ($\sim$70\% of the total).  Preliminary results from a
wide-field, ground-based $BVR$ study of NGC~3379's GC system by
Butterworth \& Hanes also indicate an excess of blue GC candidates
(D.~Hanes, private communication).
However, the $V-I$ distribution of the inner 0.5$\arcm$ of NGC~3379's
GC system from the HST study of \citet{kw01} appears fairly evenly
distributed over the range of $V-I$ colors.  In the next section we
show that there may be a color gradient in NGC~3379's GC system in the
sense that the ratio of blue to red GCs increases with radius.  In
this case it might make sense that the HST data in the inner region do
not show an excess of blue clusters but that the wide-field data do.
If the gradient and the excess of blue GCs in NGC~3379 are real, they
underscore the importance of obtaining wide-field data to deriving
accurate total numbers of red and blue GCs for comparison to galaxy
formation models (see Section~\ref{section:Es discussion}).

\subsection{Color Gradients}
\label{section:Es gradients}

\begin{figure}
\plotone{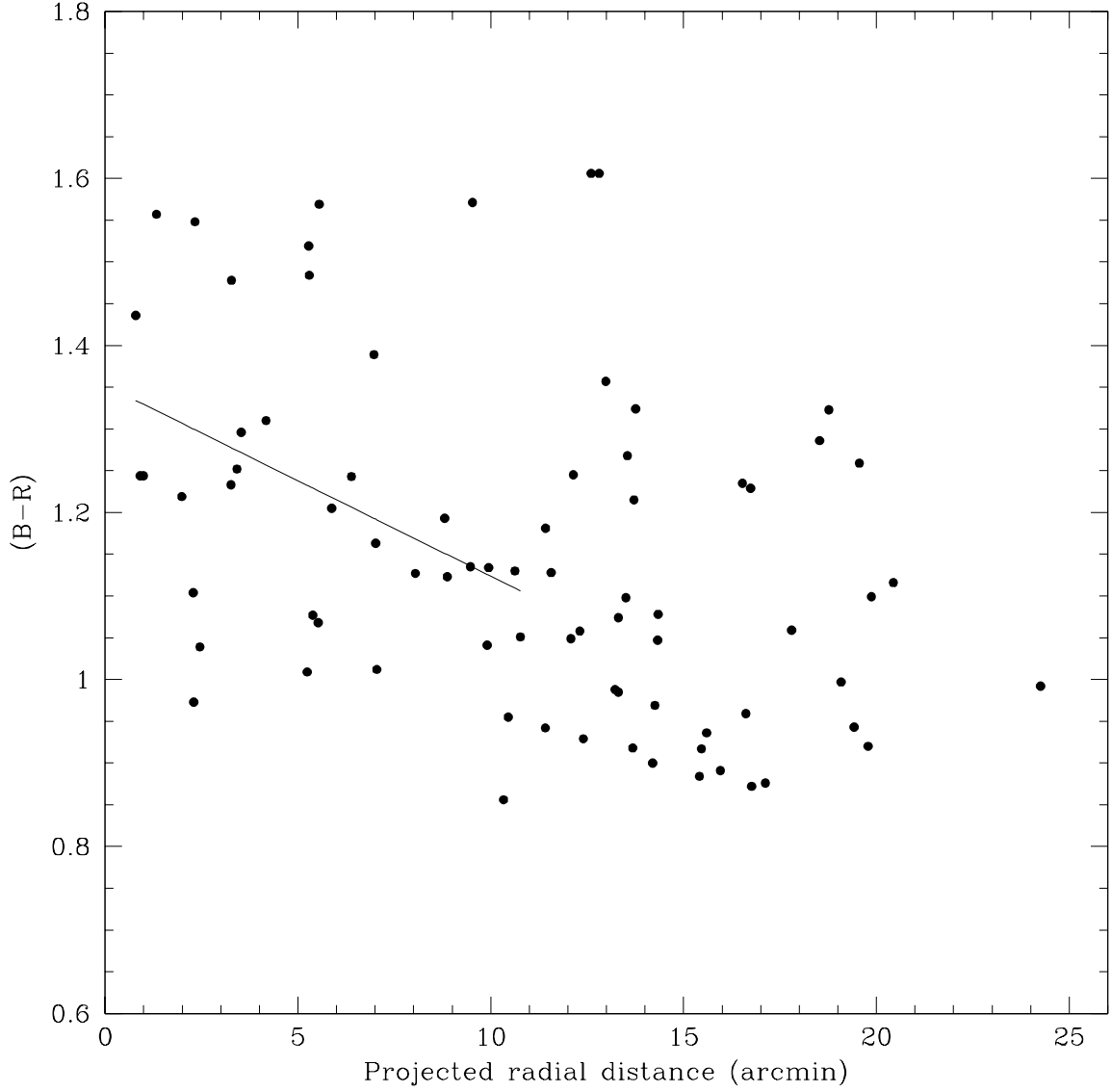}
\caption{$B-R$ color versus projected radial distance of the 90\%
sample of GC candidates in NGC~3379.  The solid line is a linear fit
to the data inside 11$\arcm$; its slope is $-$0.023$\pm$0.010.
\protect\label{fig:n3379 color gradient}}
\end{figure}

\begin{figure}
\plotone{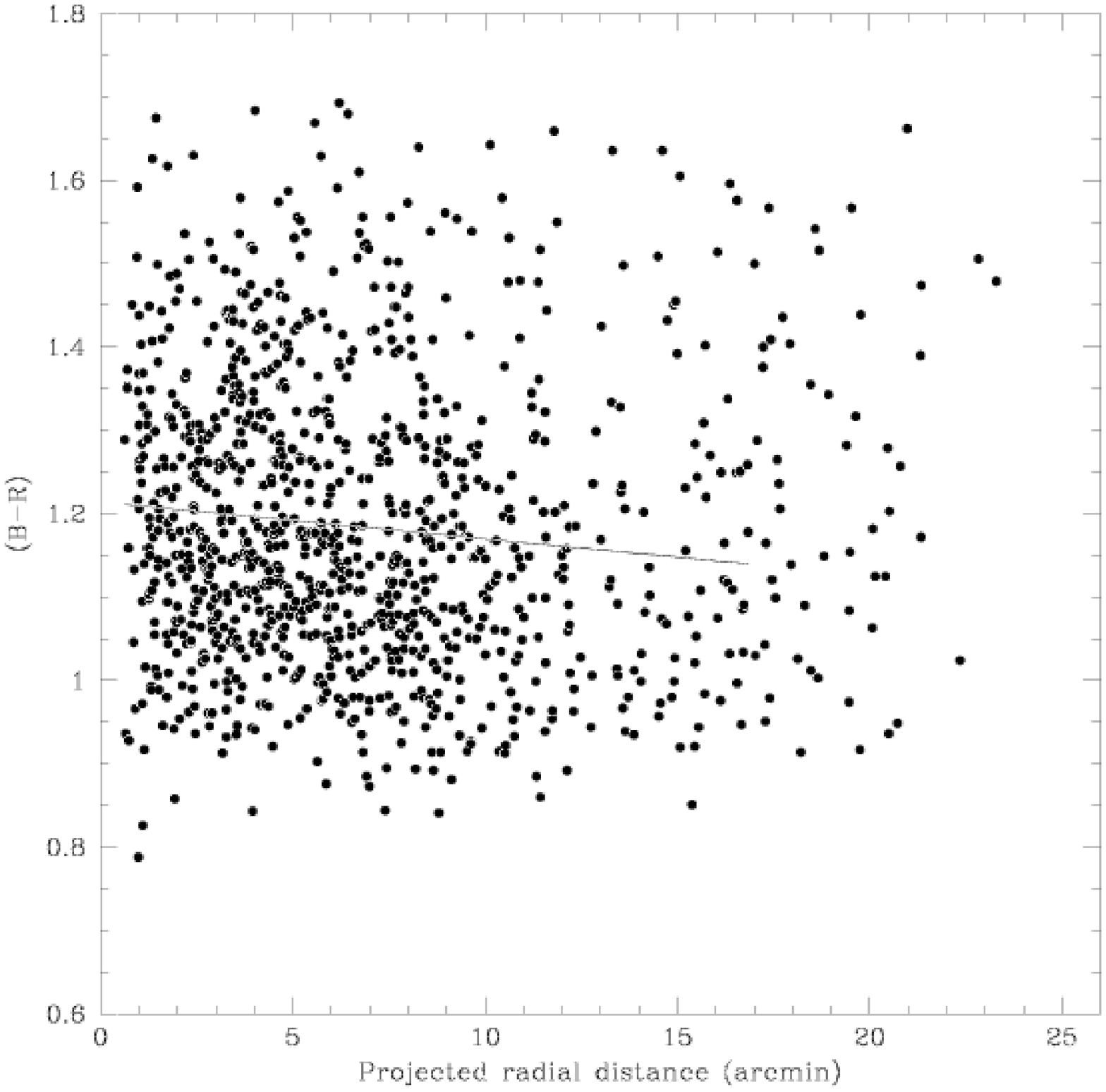}
\caption{$B-R$ color versus projected radial distance of the 90\%
sample of GC candidates in NGC~4406.  The solid line is a linear fit
to the data inside 17$\arcm$, with a slope of $-$0.004$\pm$0.001.
\protect\label{fig:n4406 color gradient}}
\end{figure}

\begin{figure}
\plotone{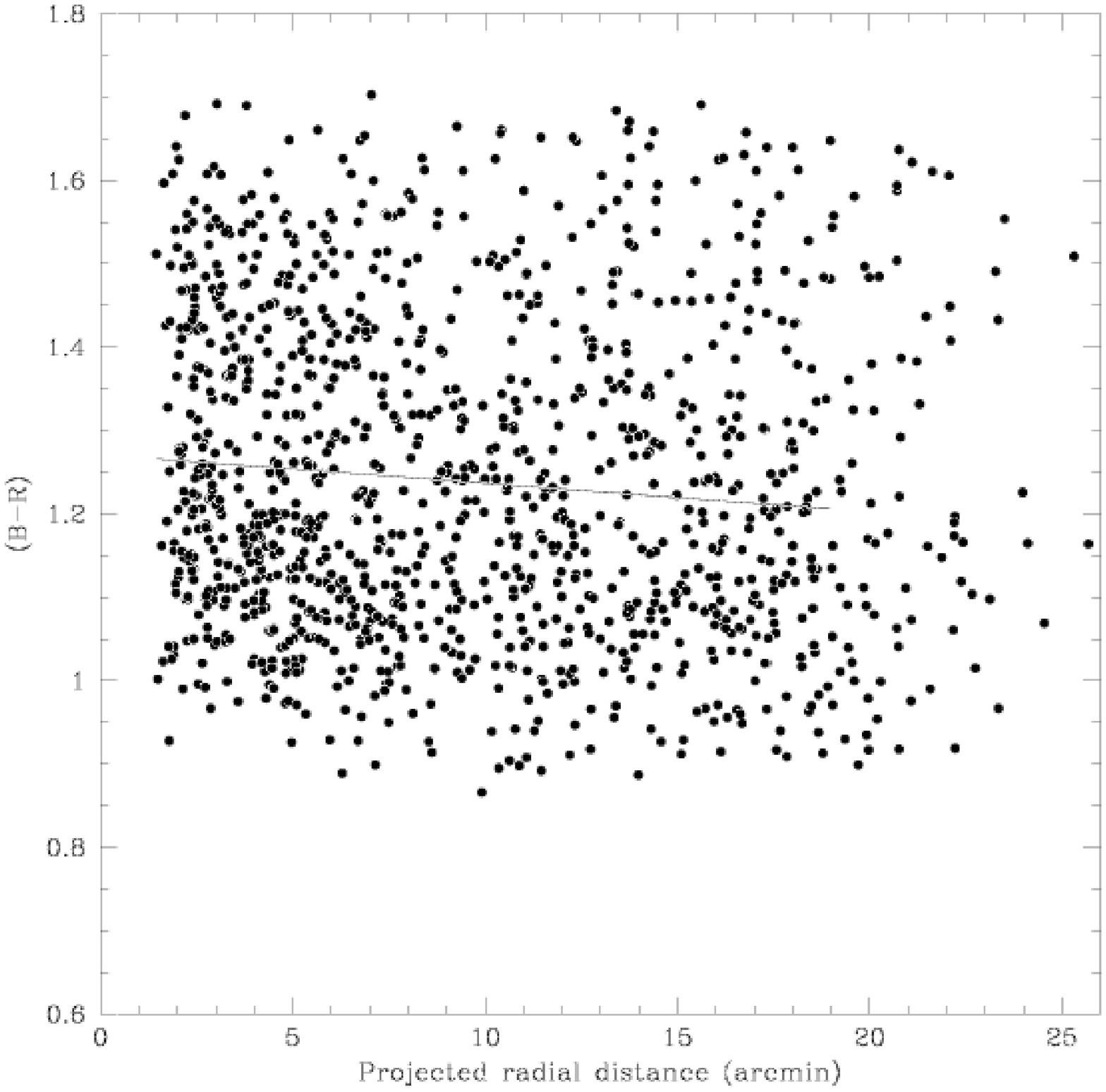}
\caption{$B-R$ color versus projected radial distance of the 90\%
sample of GC candidates in NGC~4594.  The solid line is a linear fit
to the data inside 19$\arcm$, with a slope of $-$0.003$\pm$0.001.
\protect\label{fig:n4594 color gradient}}
\end{figure}

A straightforward way to look for color gradients in each galaxy's GC
system is to plot color versus projected radius for a complete sample
of GC candidates.  Accordingly, Figures~\ref{fig:n3379 color gradient}
through \ref{fig:n4594 color gradient} show $B-R$ versus radius for the
90\% samples.  We fitted lines to the data for each galaxy, imposing
the same radial cut that was used to create the input for KMM.  The
best-fit lines are shown in the figures.  To calculate metallicity
gradients, we converted the $B-R$ values to [Fe/H] (using the
relationship derived in Paper~I) and fitted lines to [Fe/H] versus
log~$r$.

All three galaxies show evidence for modest negative color gradients
in their overall GC populations.  For NGC~4406 and NGC~4594, the
slopes of the best-fit lines in the color-radius plane are
$\Delta(B-R)/\Delta~r$ $=$ $-$0.004$\pm$0.001 and $-$0.003$\pm$0.001,
respectively.  These translate to metallicity gradients of
$\Delta$[Fe/H]/$\Delta$~log~$r$ $=$ $-$0.16$\pm$0.06 and
$-$0.19$\pm$0.06.  NGC~3379 has a formally steeper gradient, but at a
lower significance: $\Delta(B-R)/\Delta~r$ $=$ $-$0.023$\pm$0.010.
The fit includes only 36 points and the scatter is large, so the
gradient is significant only at the 2-$\sigma$ level.  If we remove
the single red GC candidate in NGC~3379 between 7$\arcm$ and
11$\arcm$, the $B-R$ gradient becomes $-$0.028+/-0.010.  Fitting
metallicity versus the log of the radius for the full (N$=$36) sample
produces $\Delta$[Fe/H]/$\Delta$~log~$r$ $=$ $-$0.60$\pm$0.31.

Two of the galaxies have had the color gradients of their GC systems
measured previously.  \citet{cohen88} studied the inner $\sim$7$\arcm$
($\sim$30~kpc) of NGC~4406's GC population with a small-format CCD
detector and $gri$ filters, and found no detectable color gradient in
this region.
Our data are consistent with this result: we find
$\Delta(B-R)/\Delta~r$ $=$ $-$0.004$\pm$0.004 inside 7$\arcm$.
\citet{bridges92} observed the inner portion of NGC~4594's GC system
with a small-format CCD and $B$ and $V$ filters. They found no
evidence for a gradient in $B-V$ with galactocentric radius for GCs in
the inner $\sim$4$\arcm$ of the galaxy. We likewise do not find a
significant color gradient in our data for this inner region.  In a
later study, Forbes, Grillmair, \& Smith (1997b) observed NGC~4594
over a larger radial range.  Using a CCD with 1.84$\arcs$-pixels and
$B$ and $I$ filters, they found no color gradient in a sample of
$\sim$400 GC candidates over the radial range 4$-$50~kpc.  The very
coarse resolution of their study (with a pixel scale $\sim$7 times
larger than the current work) may have resulted in significant
contamination from background galaxies, which would affect their
ability to detect a gradient.

A cautionary note is that although we have carefully minimized
contamination in the GC samples through multi-color selection and
image analysis, the proportion of GCs relative to contaminating
objects inevitably decreases with increasing radius. The gradients we
find may be affected by this if the contaminating objects are strongly
biased in color relative to the GCs. To investigate this possibility,
we examined the $B-R$ color distributions of the GC candidates with
radial distances beyond the apparent extent of the each galaxies' GC
system.  In NGC~4594 and NGC~3379, the blue and red proportions of
these likely contaminants are very similar to the proportions we
derived for the GC systems.  In NGC~4406, the contaminants are
uniformly distributed in $B-R$.  What this indicates is that the
contaminating objects are not likely to be heavily biased toward the
blue or red with respect to the GCs, which seems to suggest that our
overall results for the color gradients are valid.

\subsection{Total Numbers and Specific Frequencies}
\label{section:Es spec freq}

The total number of GCs in the three target galaxies can be calculated
by integrating the best-fit deVaucouleurs profile from $r$ $=$ 0 to a
specified outer radius.  We have chosen to stop the integration for
each galaxy at the point at which the GC surface density becomes
consistent with zero and remains so until the end of the data. As
discussed in Section~\ref{section:Es radial profiles} this occurs at
11$\arcm$, 17$\arcm$ and 19$\arcm$ for NGC~3379, NGC~4406, and
NGC~4594, respectively.  To facilitate comparison between galaxies,
the total number of GCs ($N_{GC}$) can be normalized by the galaxy
luminosity or mass; the luminosity-normalized specific frequency,
$S_N$, is defined as
\begin{equation}
{S_N \equiv {N_{GC}}10^{+0.4({M_V}+15)}}
\end{equation}
\citep{hvdb81}
and the mass-normalized number of GCs, $T$, is
\begin{equation}
T \equiv \frac{N_{GC}}{M_G/10^9\ {\rm M_{\sun}}}
\end{equation}
\citep{za93}, where $M_G$ is the stellar mass of the host galaxy.
Table~\ref{table:Es S values} lists the total numbers and specific
frequencies derived for the three target galaxies.  Note that to
calculate $T$, we have assumed mass-to-light ratios ($M/L_V$) of 10
for the two ellipticals and $M/L_V$ $=$ 7.6 for the S0 NGC~4594,
following \citet{za93}. Table~\ref{table:Es S values} also includes the
radial extent of each GC system (derived by converting the integration
limit used to calculate $S_N$ into a physical distance) along with a
list of $S_N$ values from previous studies.  Note that to allow
straightforward comparison between our $S_N$ values and others', we
have normalized $N_{GC}$ from the previous studies with the same $M_V$
values as used in this work; thus any differences in $S_N$ reflect
differences in total numbers rather than in host galaxy magnitude or
distance.

To estimate the error associated with our $S_N$ values, we calculated
the Poisson errors on the total numbers of GCs and contaminating
objects, uncertainties related to fitting the GCLF, and uncertainty in
the total magnitude of the galaxy.  For the latter, we assumed that
the galaxy magnitudes could be brighter or fainter by as much as three
times the error on $V_T^0$ given in RC3 \citep{devauc91}.  We added
each of these uncertainties in quadrature to produce an initial
estimate of the error.  Next, we investigated how the choice of outer
integration limit for the radial profile affects $S_N$.  In each
galaxy's radial distribution, the GC surface density becomes negative
in one or more of the outer annuli.  For this test, we calculated new
$S_N$ values by integrating the best-fit deVaucouleurs profiles out to
the last bin before $\sigma_{\rm GC}$ becomes negative.  For NGC~4406
and NGC~4594, the integration limit became 21$\arcm$ and 22$\arcm$,
respectively.  NGC~3379's profile is somewhat noisier and $\sigma_{\rm
GC}$ is negative by 11$\arcm$, which is where we had originally
stopped the integration.  In this case we continued the integration to
15$\arcm$, since $\sigma_{\rm GC}$ is negative in the next two annuli
after that.  The new $S_N$ values for NGC~4406 and NGC~4594 matched
the original values within our estimated errors.  $S_N$ for NGC~3379
increased by 0.27, whereas we had estimated an initial error of 0.2;
to account for this, the final $S_N$ for NGC~3379 is listed in
Table~\ref{table:Es S values} as 1.2$\pm$0.3.

As a last check on the derived $S_N$ values, we investigated whether
summing the actual data points in the radial distributions rather than
integrating the fitted profiles produces $S_N$ values consistent with
the final values and errors.  For NGC~3379 and NGC~4594, $S_N$ from
summing the data matched the values in Table~\ref{table:Es S values}.
Because NGC~4406's GC distribution drops below the deVaucouleurs law
well before the chosen integration limit, $S_N$ from summing the
points is 0.1 smaller than the lower limit given in the table.

\subsection{Mass-Normalized Number of Blue Globular Clusters}
\label{section:Es Tblue}

The relative numbers of blue, metal-poor GCs in the massive
ellipticals and spirals we have targeted are especially relevant to
the question of how such galaxies formed.  The AZ92 merger model, for
example, predicts that spirals and ellipticals should have similar
numbers of metal-poor GCs, since in their scenario, the metal-poor GC
populations in ellipticals come directly from the progenitor spirals.
In hierarchical structure formation scenarios like that of
\citet{santos03}, in which metal-poor GC formation occurs over a
finite period in the early Universe, massive galaxies in dense
environments naturally have larger proportions of metal-poor GCs than
lower-mass field galaxies.  Accurate determinations of how many blue,
metal-poor GCs exist around a given galaxy are therefore important for
testing the predictions and assumptions of these models.

Section~\ref{section:Es color distributions} described how KMM was used
to estimate the proportion of blue GCs in the three target galaxies.
For NGC~3379, we estimated that $\sim$70\% of the GC population is
blue. For the other two galaxies, the proportion varied by a few
percent depending on the conditions used to fit the data with KMM, so
we have taken the average of the results as the final value.  For
NGC~4406, the fraction is .62 and for NGC~4594 it is .63.  Combining
these values with the number of GCs normalized by galaxy mass yields
$T_{\rm blue}$ for each galaxy; the results are listed in column 7 of
Table~\ref{table:Es S values}.

\section{Discussion}
\label{section:Es discussion}

The main result of this paper is that by utilizing wide-field CCD
detectors and techniques to both reduce and quantify contamination
from non-GCs, we have been able to derive robust global values for the
GC system properties of the early-type galaxies NGC~3379, NGC~4406,
and NGC~4594.  Here we discuss our findings as well as examine their
implications for the formation of galaxies and their GC populations.
As explained in the Introduction, several models for the formation of
early-type galaxies and their GC systems have been proposed; these
include spiral-spiral mergers (AZ92), multiple phases of star
formation (FBG97), dissipational collapse with accretion (CMW98), and
hierarchical merging (e.g., Beasley et al.\ 2002, Santos 2003).

{\it Global specific frequency.---} Table~\ref{table:Es S values} shows
that the $S_N$ value of the giant elliptical NGC~4406 is reduced by
$\sim$20\% compared to previous values.
%
%
The $S_N$ value for the Virgo elliptical NGC~4472 (from Paper~I) is
also $\sim$20\% smaller compared to past work.  Both values agree with
the previous determinations within their quoted errors, which were
about twice as large as ours.  The $S_N$ values for NGC~3379 and
NGC~4594 are nearly identical to the values found in past studies, but
their uncertainties have been reduced to one-third to one-half their
former size.

The previous determinations of $S_N$ for NGC~4406 and NGC~4472 were
done using photographic plates and it is possible that the numbers
were undercorrected for contamination.  The power-law profile fits
that were typically used to calculate total numbers may have also
contributed to the discrepancy with our values.  These galaxies' GC
systems are more extended than those of the less luminous galaxies in
the sample, so overestimating the GC counts at large radius has more
of an effect because the summation to derive $N_{\rm GC}$ is done over
a larger area.

An interesting and potentially meaningful result is that (despite
$S_N$ and its uncertainties having been reduced) the two giant cluster
ellipticals in our sample still match closely in terms of the number
of GCs per unit luminosity or mass.  Their $T_{\rm blue}$ values are
also similar.  NGC~3379 has $S_N$~$\sim$~1, the lowest value in the
early-type sample.  This is comparable to $S_N$ for spirals of similar
luminosity (see Paper~II).  NGC~4594's $S_N$ value is midway between
that of the cluster ellipticals and spirals, which perhaps makes sense
since morphologically, it falls somewhere between spirals and
ellipticals.  On the other hand, NGC~4594 is a luminous galaxy and has
an absolute magnitude like that of NGC~4406.  When examined as a
whole, the results from the early-type galaxy sample suggest that
$S_N$ is dependent at least in part on luminosity/mass of the host
galaxy, with possible additional effects from morphology and/or
environment.

{\it Specific frequency of blue GCs.---} As noted above, the
mass-normalized numbers of blue GCs ($T_{\rm blue}$) for the two
luminous Virgo cluster ellipticals in the sample are essentially the
same, at 2.6 for NGC~4472 and 2.5 for NGC~4406.  NGC~4594, which is of
similar luminosity to NGC~4406 and is an early-type field galaxy, has
$T_{\rm blue}$~$=$~2.0. The moderate-luminosity field elliptical
NGC~3379 has the smallest $T_{\rm blue}$ of the sample, 1.0.

In Paper~II, we estimated $T_{\rm blue}$ for the Sab spiral NGC~7814,
a field galaxy with $M^T_V$ $=$ $-$20.4.  Its $T_{\rm blue}$ value is
in the range 1$-$2, so somewhere between that of NGC~3379 and
NGC~4594.  The Milky Way and M31 have absolute $V$ magnitudes of
$-$21.3 and $-$21.8 and $T_{\rm blue}$ $\sim$ 0.9 and 1.2,
respectively (see discussion in Paper~II).  The $T_{\rm blue}$ values
of these spirals (all of them located in low-density environments) are
on the whole comparable to that of NGC~3379, and smaller than $T_{\rm
blue}$ for the giant cluster Es.

The above comparison, taken at face value, suggests that it is
unlikely that luminous cluster ellipticals like NGC~4406 and NGC~4472
could have formed from the merger of spirals like the Milky Way, M31,
or NGC~7814 in the way that AZ92 envisioned.  There do not appear to
be enough blue, metal-poor GCs in typical spirals that we see today
(assuming the three we are using as examples are typical) to account
for the large metal-poor populations of GCs in luminous, high-$S_N$
ellipticals.  It {\it does} seem possible, however, that the blue GC
population in a more moderate-luminosity elliptical like NGC~3379
could have come from the merger of spirals like the Milky Way and
NGC~7814, since $T_{\rm blue}$ for these galaxies is similar.  $T_{\rm
blue}$ for NGC~4594 is larger than that for the Galaxy and M31 but
falls at the high end of the range we estimate for NGC~7814, so we
cannot rule out that an AZ92-like merger could account for its
metal-poor GC system.

It is relevant to note again at this point that when calculating $T$
for a given galaxy, one adopts a mass-to-light ratio to convert $M_V$
to total stellar mass.  Following the convention set in the paper that
introduced the $T$ parameter \citep{za93}, we used the same $M/L_V$
for all three ellipticals.  In fact, elliptical galaxy mass-to-light
ratios almost certainly vary with luminosity. The simplest reason for
this is that elliptical galaxies follow a color-magnitude relation.
More luminous ellipticals are redder, and these redder stellar
populations are associated with slightly higher $M/L_V$.  Assuming
that the color-magnitude relationship is caused by metallicity effects
leads to $M/L_V$~$\propto$~$L^{0.07}$ \citep{dressler87}, whereas
assuming that age is the primary contributor yields a slightly larger
exponent ($\sim$0.10; see, e.g., Zepf \& Silk 1996). A steeper
dependence of $M/L_V$ on $L$ is suggested by studies of the
fundamental plane (e.g., Dressler et al. 1987; Faber et al. 1987;
Kormendy \& Djorgovski 1989).  However, this difference is typically
ascribed to systematic breaking of the assumption of homology along
the elliptical galaxy sequence, or by a larger dark matter
contribution in more luminous ellipticals (see, e.g., Pahre,
Djorgovski, \& de~Carvalho 1995 and references therein). Neither of
these effects reflect differences in stellar mass-to-light ratios, so
the appropriate relation to use to examine the effect of changing
$M/L_V$ on our results is that from the stellar populations
differences, $M/L_V$~$\propto$~$L^{0.10}$.

Taking into account that the mass-to-light ratios of ellipticals may
vary as $L^{0.10}$ reduces the $T_{\rm blue}$ values for the more
luminous ellipticals, NGC~4406 and NGC~4472, by a factor of 1.1$-$1.2
relative to the value for NGC~3379.
%
%
Our observed values of $T_{\rm blue}$ for NGC~4472 and NGC~4406 are
2.5 times that for NGC~3379, and $\sim$2 times larger than the mean
$T_{\rm blue}$ value for the three spirals mentioned above.  Therefore
(as we noted in Paper~II), using variable instead of constant
mass-to-light ratios accounts for only half or less of the apparent
difference in $T_{\rm blue}$ for ellipticals of different luminosity.
Thus the differences in $T_{\rm blue}$ between the luminous
ellipticals and the less luminous galaxies in the sample appear to be
real.

An assumption we make when comparing $T_{\rm blue}$ for spirals and
ellipticals {\it at the present day} is that dynamical destruction has
not substantially changed the numbers of metal-poor GCs since the time
that the spiral-spiral mergers might have occurred, or that it has had
a fairly equal effect on the halo GC populations in both types of
galaxies.  As we discussed in Paper~II, it is possible that this
assumption is incorrect and that more GC destruction may have taken
place in lower-luminosity ellipticals and spirals compared to
high-luminosity ellipticals.  If this is the case, it could
potentially explain the observed differences in $T_{\rm blue}$ for
typical-luminosity ellipticals and spirals versus high-luminosity
ellipticals.  Avenues for further work in this area include developing
new observational tests of the role of dynamical destruction, as well
as modeling GC destruction in galaxies with a wider range of
properties (e.g., spirals of varying luminosity).

In Paper II we also discussed the idea that hierarchical merging
scenarios for the origins of spirals and ellipticals may be able to
explain the variation of $T_{\rm blue}$ with galaxy luminosity and
environment.  In this type of picture, metal-poor GC formation occurs
at high redshift in protogalactic fragments or ``building blocks''
that merge to form larger structures.  In the \citet{santos03}
hierarchical model, GC and structure formation are temporarily
suppressed when reionization occurs.  The protogalactic fragments in
higher-density regions (e.g., in locations that eventually become
galaxy clusters) collapse and begin forming metal-poor GCs earlier
than those located in lower-density regions.  As a result, massive
galaxies in high-density environments (like giant cluster ellipticals)
have the largest numbers of metal-poor GCs per unit luminosity or
mass.  Less massive galaxies in poorer environments (like spirals and
ellipticals in the field) naturally have smaller $T_{\rm blue}$
values.  To complete the scenario, stellar evolution enriches the
intergalactic medium during the interval when GC formation is
suppressed, so that GCs formed after this interval are comparatively
metal-rich.  Gaseous merging and hierarchical assembly continue,
triggering additional metal-rich GC formation and eventually resulting
in massive galaxies that have GC systems with both metal-poor and
metal-rich subpopulations.  Hierarchical merging --- with GC formation
being temporarily suppressed (perhaps by reionization) --- thus seems
to provide a sensible overall framework with which to understand the
observation that $T_{\rm blue}$ is apparently larger for the luminous
cluster ellipticals and smaller for the field galaxies in the survey.

{\it Color distributions.---} The $B-R$ color distributions for the
galaxies in the early-type sample are better fit by two Gaussians than
a single one at 99.8\% confidence or higher.  (The distribution for
NGC~4406 may actually show hints of three peaks.)  The blue peaks of
the color distributions are centered at $B-R$ $\sim$ 1.1, perhaps
suggesting a similar origin for the metal-poor GC populations.  The
locations of the red (metal-rich) peaks vary slightly, ranging from
$B-R$ $\sim$ 1.3 to 1.5.  The global ratios of red to blue GCs in the
two Virgo ellipticals and NGC~4594 are roughly similar, at
$\sim$0.60$-$0.66.  NGC~3379 appears to have a larger proportion of
blue GCs and its red-to-blue ratio is $\sim$0.40.
%

Different models produce different expectations for the proportions of
red and blue GCs in elliptical systems. 
CMW98 state that giant ellipticals with high GC specific frequencies
should have an excess of metal-poor GCs, 
because these galaxies will have captured (through accretion or tidal
stripping) proportionately larger numbers of metal-poor clusters in
their outer regions.  Similarly, FBG97 predict a trend in the sense
that ellipticals with increasing specific frequency will have larger
proportions of metal-poor GCs, and low-luminosity, low-$S_N$
ellipticals should have relatively more metal-rich than metal-poor
GCs.
At least for the galaxies in our sample, we see the opposite behavior.
The two cluster ellipticals with the highest $S_N$ values show the
same ratio of metal-rich to metal-poor GCs, and the same ratio exists
in the S0 galaxy with moderate $S_N$.  It is the lowest-luminosity,
lowest-$S_N$ galaxy in the sample, NGC~3379, that appears to have a
larger proportion of blue GCs.

A hierarchical merging scenario may make more sense in terms of our
results for the color distributions and relative proportions of red
and blue GCs.  In this type of picture, the massive cluster
ellipticals might be expected to have a higher proportion of
metal-rich GCs because they generally experience a larger number of
the gas-rich major mergers over their histories than less massive
galaxies do.  As part of their simulations of hierarchical galaxy
formation, \citet{beasley02} produce broadband color distributions for
the GC systems of galaxies of different luminosities.  The color
distributions show a variety of morphologies, from distinctly bimodal
(with either red or blue peaks dominating) to fairly flattened with
only a small dip between the populations.
A number of them
appear at least qualitatively very much like our observed
distributions, which likewise show a range in appearance
and which presumably reflect their different evolutionary histories.

{\it Color gradients.---} NGC~4406 and NGC~4594 both exhibit small but
significant (at 3-$\sigma$ or better) gradients in their overall GC
populations, in the sense that the ratio of blue to red GCs increases
slightly with increasing galactocentric radius.  This is in contrast
to NGC~4472, which shows a weak color gradient in its inner regions
but none in its overall population (Paper~I).  The $B-R$ versus radius
data for NGC~3379 show a relatively steep gradient but the
uncertainties are large (again possibly due to small-number
statistics) and it is significant at only about the 2-$\sigma$ level.

Color gradients in the GC systems of early-type galaxies are possible
or expected in all the formation scenarios we are considering.  In
AZ92, FBG97, and the scenarios involving hierarchical merging, the
blue, metal-poor GCs are expected to have a more extended distribution
than the red GCs because the latter are younger and formed from gas
that experienced subsequent dissipation.  In the CMW98 model, the
metal-poor GCs should be more extended relative to the metal-rich
population, at least in giant ellipticals located near the centers of
clusters, because large numbers of them are accreted into the
galaxies' outer regions either through capture of dwarf galaxies or
tidal stripping.

A specific prediction of FBG97 concerning color/metallicity gradients
is that galaxies with larger specific frequencies should have steeper
radial metallicity gradients, due to the greater numbers of metal-poor
GCs at large radius.  Our data appear to be inconsistent with this
prediction, since NGC~4472, NGC~4406 and NGC~4594 show either no
radial gradient or a very shallow one, and NGC~3379 appears to have
the steepest color gradient of the sample.

It is worth noting that our measured color distributions and color
gradients are global results and essentially cover the full radial
extent of their GC systems.  It is important to observationally test
the models using these global properties since (as we saw in NGC~4472)
color gradients can be present in the inner regions of galaxies but
not there when one takes into account the entire system.  Moreover,
some of the models (e.g., CMW98, Santos 2003) make specific
predictions with regard to GC populations in the outermost regions of
ellipticals and good-quality, wide-field data are what is needed to
address these.

{\it Radial Distributions.---} The best-fitting deVaucouleurs laws and
power laws to the radial distributions of all three ellipticals in the
sample have slopes that are the same within the errors ($-$1.6 for the
deVaucouleurs law and $-$1.2 to $-$1.4 for the power law).  The S0
galaxy NGC~4594 has the steepest slope ($-$2.1 for the deVaucouleurs
law and $-$1.9 for the power law).  The radial profiles for three of
the four galaxies are better fit by deVaucouleurs laws than power
laws, which typically overestimate the GC surface density by a larger
amount in the outer regions of the profile.  (For NGC~3379, the
deVaucouleurs law and power law both provide good fits.)

Tidal stripping of GCs is thought to play a part in the evolution of
the GC populations of galaxies in clusters, and is explicitly included
in the FBG97 and CMW98 formation pictures.  Our results for the GC
radial distributions of two of the galaxies in the sample may have
some significance with regard to tidal stripping and its role in the
evolution of GC systems.  NGC~3379 has the smallest apparent extent
($\sim$30~kpc) and the lowest $S_N$ of the early-type sample.  Since
it is the largest galaxy in its environment, it is not likely to have
lost a substantial number of its GCs due to tidal stripping. Thus its
low specific frequency appears to be an intrinsic property of this
galaxy.  Finally, FBG97 predict that galaxies that have been
tidally-stripped should have lower-than-average $S_N$ values.
However, the one galaxy in our sample that shows a hint that its GC system
might be tidally truncated, NGC~4406, has a GC specific frequency that
is almost identical to that of NGC~4472, which is a luminous giant
elliptical located near the center of the Virgo cluster.

\section{Summary}
\label{section:Es summary}

As part of a wide-field CCD survey of the GC populations of a sample
of normal galaxies, we have investigated the GC system properties of
the early-type galaxies NGC~3379, NGC~4406, and NGC~4594. Below is a
summary of our findings.

1. The radial coverage of the data extends to $\sim$24$-$25$\arcm$, which
corresponds to $\sim$70$-$100~kpc at the distances of our targets.
For all three galaxies, the GC surface density decreases until it is
consistent with zero well before the end of the radial profile,
strongly suggesting that we have observed the entire extent of the GC
systems.  The physical radius at which the GC surface density becomes
consistent with zero ranges from 30~kpc for the least luminous elliptical
to 80$-$100~kpc for the two more luminous galaxies.
%

2. The $B-R$ color distributions of the GCs of the target galaxies are
   better fit by a bimodal distribution than a unimodal one at
   $\geq$99.8\% confidence.  The blue GC distributions are centered at
   $B-R$~$\sim$~1.1 in all three galaxies; the metal-rich populations
   are centered between $B-R$ $\sim$1.3 and 1.5.  The two more
   luminous galaxies, NGC~4406 and NGC~4594, have $\sim$60\% blue and
   $\sim$40\% red GCs.  NGC~4472 shows similar proportions (Paper~I).
   In the lower-luminosity elliptical, NGC~3379, blue GCs make up
   $\sim$70\% of the total population.

3. All three galaxies presented here show modest negative color
   gradients in their GC systems, caused by the increasing ratio of
   blue to red GCs with increasing projected radial distance.

4.  $S_N$ for NGC~4406's GC population is 3.5$\pm$0.5, which
represents a $\sim$20\% reduction in the number of GCs compared to
previous estimates.  We obtained a similar result for the other giant
Virgo elliptical in the sample, NGC~4472 (Paper~I).  NGC~3379 and
NGC~4594 have $S_N$ of 1.2$\pm$0.3 and 2.1$\pm$0.3, respectively;
these values are similar to those from past studies but have much
smaller errors.

5. The mass-normalized numbers of blue, metal-poor GCs ($T_{\rm
blue}$) in the Virgo cluster ellipticals are almost identical and are
$\sim$2.5 times larger than the value for the field elliptical
NGC~3379, whose blue GC population is comparable to that of spirals of
similar mass.  The field S0 NGC~4594 has $T_{\rm blue}$ between that
of NGC~3379 and the cluster ellipticals.  To date, our survey data
suggest that merging the blue GC populations of typical spirals is not
likely to produce enough metal-poor GCs to account for the blue
populations in luminous, high-$S_N$ cluster ellipticals; however, it
may be able to account for the blue GC populations of ellipticals of
more moderate luminosity.  The proportion of blue GCs is roughly
constant
for three of the galaxies in the early-type sample, and slightly
larger
for the galaxy with the lowest specific frequency.  This result is not
consistent with galaxy formation models (e.g., FBG97 and CMW98) that
predict that galaxies with larger $S_N$ should have proportionately
more metal-poor GCs.

6. We compare our results with the predictions and assumptions of a
   number of models for the formation of massive galaxies and their GC
   systems.  The general scenario that appears most consistent with
   the observations is one in which metal-poor GCs form in the early
   Universe in merging protogalactic building blocks.  Subsequent
   hierarchical merging produces metal-rich GCs and eventually results
   in today's massive galaxies.  Detailed models within this
   hierarchical framework are currently being produced and our data on
   the global properties of massive galaxies' GC systems are likely to
   prove valuable for helping to constrain and test the theoretical
   work in this area.



\acknowledgments

K.L.R. gratefully acknowledges financial support from a NASA Graduate
Student Researchers Fellowship for this project. S.E.Z. acknowledges
support for this work from the Michigan State University Foundation
and NASA Long-Term Space Astrophysics Grant NAG5-11319.  We thank
Arunav Kundu and John Salzer for obtaining post-calibration data at
the WIYN telescope in 2001 April and 2002 March, respectively.  We
have benefited from discussions with Kathryn Johnston and Brad
Whitmore, which resulted in improvements to the paper.  We also thank
the referee, Terry Bridges, for providing valuable suggestions and
comments on the manuscript.  Finally, we are grateful to the staff at
Kitt Peak National Observatory and WIYN for assistance during the
observing runs.  This research has made use of the NASA/IPAC
Extragalactic Database (NED), which is operated by the Jet Propulsion
Laboratory, California Institute of Technology, under contract with
the National Aeronautics and Space Administration.


\clearpage

%
%

\clearpage

\clearpage

\clearpage
\begin{deluxetable}{lccrrl}
\tablecaption{Basic Properties of Galaxies in the Early-type Sample}
\tablewidth{370pt}
\tablehead{\colhead{Name}
&\colhead{Type}
&\colhead{$m-M$} &\colhead{Dist} &\colhead{$M_V^T$} &\colhead{Environment}\\
\colhead{} & \colhead{} & \colhead{} & \colhead{(Mpc)} & \colhead{} & \colhead{}}
\startdata
NGC4472 (M49)& E2 & 31.12 & 16.7 & $-$23.1 & Virgo cluster\\ 
NGC4406 (M86)& E3 & 31.12 & 16.7 & $-$22.3 & Virgo cluster\\ 
NGC3379 (M105)& E1 & 30.12 & 10.6 & $-$20.9 & Leo-I group\\ 
NGC4594 (M104)& S0 & 29.95 & 9.8 & $-$22.4 & Field\\ 
\enddata
\tablecomments{Virgo cluster distance is from Whitmore et al.\ (1995)
(HST observations of M87's GCLF); other distances are from Tonry et
al.\ (2001) (surface brightness fluctuations).  Magnitude for NGC~4472
from Rhode \& Zepf (2001); all others from combining $V_T^0$ from RC3
(deVaucouleurs et al.\ 1991) with $m-M$.}
\protect\label{table:Es properties}
\end{deluxetable}

\begin{deluxetable}{lrrr}
\tablecaption{Aperture Corrections Used for Photometry of Mosaic Sources}
\tablewidth{310pt}
\tablehead{\colhead{} & \colhead{NGC~3379}& \colhead{NGC~4406}& \colhead{NGC~4594}}
\startdata
$B......$ & $-$0.149 $\pm$ 0.003 & $-$0.167 $\pm$ 0.005 & $-$0.138 $\pm$ 0.002\\
$V......$ & $-$0.171 $\pm$ 0.004 & $-$0.224 $\pm$ 0.006 & $-$0.262 $\pm$ 0.004\\
$R......$ & $-$0.237 $\pm$ 0.006 & $-$0.159 $\pm$ 0.005 & $-$0.202 $\pm$ 0.004\\
\enddata
\protect\label{table:Es aper corr}
\end{deluxetable}

\begin{deluxetable}{lccc}
\tablecaption{Extinction Corrections Used for Photometry of Mosaic Sources}
\tablewidth{250pt}
\tablehead{\colhead{} & \colhead{NGC~3379}& \colhead{NGC~4406}& \colhead{NGC~4594}}
\startdata
$A_B......$ & 0.105 & 0.126 & 0.222\\
$A_V......$ & 0.081 & 0.097 & 0.171\\
$A_R......$ & 0.065 & 0.077 & 0.138\\
\enddata
\protect\label{table:Es ext corr}
\end{deluxetable}

\begin{deluxetable}{lrrr}
\tablecaption{50\% Completeness Limits of Mosaic Images}
\tablewidth{250pt}
\tablehead{\colhead{} & \colhead{NGC~3379}& \colhead{NGC~4406}& \colhead{NGC~4594}}
\startdata
$B......$ & 23.56 & 25.44 & 24.63\\
$V......$ & 24.09 & 24.17 & 23.98\\
$R......$ & 23.40 & 23.48 & 23.64\\
\enddata
\protect\label{table:Es completeness}
\end{deluxetable}

\begin{deluxetable}{lllrc}
\tablecaption{HST WFPC2 Observations Analyzed for this Study}
\tablewidth{350pt}
\tablehead{\colhead{Proposal ID} & \colhead{Target Name}& \colhead{PI}&
\colhead{Ang Sep} & \colhead{Filter}\\
\colhead{} & \colhead{} & \colhead{} & \colhead{($\arcm$)} & \colhead{}}
\startdata
NGC 3379: & & & & \\
5512    &   NGC 3379-Nuc1 &	Faber	 & 0.5&	F814W\\
7909    &   Any		 &	Casertano& 3.0&	F606W\\
8059    &   List-2	 &	Casertano& 5.2&	F606W\\
5233    &   NGC 3379-Pos3 &	Westpfahl& 6.2&	F814W\\
5512    &   NGC 3384	 &	Faber	 & 6.9&	F814W\\
\cr
\tableline					
\cr
NGC 4406: & & & & \\
5512	&	NGC 4406	&	Faber 	 & 0.4 &	F814W\\
7377	&	VCC 896	&	Miller	 & 9.4 &	F555W\\
7909	&	Any	&	Casertano& 10.0&	F606W\\
7202	&	Parallel Field&	Windhorst& 10.1&	F814W\\
7566	&	NGC 4374	&	Green	 & 13.5& F606W\\
\cr
\tableline					
\cr
NGC 4594: & & & & \\
5512&	NGC 4594-Nuc1  &	Faber	 & 0.5&	F814W\\
5091&	Parallel Field&	Groth	 & 4.5&	F814W\\
5369&	Hi Lat	      &	Griffiths& 8.2&	F814W\\
\enddata
\protect\label{table:Es hst data}
\end{deluxetable}

\begin{deluxetable}{rrrrrrrrr}
\tablewidth{0pt}
\tablecaption{Corrected Radial Profiles of GCs in Target Galaxies}
\tablehead{
\multicolumn{3}{c}{NGC~3379} &
\multicolumn{3}{c}{NGC~4406} &
\multicolumn{3}{c}{NGC~4594} \\
\cline{1-3} \cline{4-6} \cline{7-9}\\
\colhead{$r$} & \colhead{$\sigma$} & \colhead{Cvg}  &
\colhead{$r$} & \colhead{$\sigma$} & \colhead{Cvg} &
\colhead{$r$} & \colhead{$\sigma$} & \colhead{Cvg}\\
\colhead{($\arcm$)} & \colhead{(arcmin$^{-2}$)} & \colhead{} &
\colhead{($\arcm$)} & \colhead{(arcmin$^{-2}$)} & \colhead{} &
\colhead{($\arcm$)} & \colhead{(arcmin$^{-2}$)} & \colhead{}
}
\startdata
1.2..... &  6.27 $\pm$  1.73 & 0.77 & 1.1..... & 24.79 $\pm$  2.95 & 0.92 & 1.4..... & 22.48 $\pm$ 16.36 & 0.02 \\ 
2.0..... &  2.52 $\pm$  0.73 & 1.00 & 2.0..... & 17.94 $\pm$  1.72 & 1.00 & 2.2..... & 15.03 $\pm$  1.57 & 0.61 \\ 
3.0..... &  1.54 $\pm$  0.50 & 1.00 & 3.0..... & 13.34 $\pm$  1.22 & 1.00 & 3.0..... &  9.26 $\pm$  0.85 & 0.90 \\ 
4.0..... &  0.54 $\pm$  0.34 & 1.00 & 4.0..... & 10.26 $\pm$  0.95 & 1.00 & 4.0..... &  5.92 $\pm$  0.58 & 0.98 \\ 
5.0..... &  0.55 $\pm$  0.32 & 0.93 & 5.0..... &  7.85 $\pm$  0.76 & 1.00 & 5.0..... &  4.87 $\pm$  0.48 & 1.00 \\ 
6.0..... &  0.43 $\pm$  0.29 & 0.89 & 6.0..... &  5.88 $\pm$  0.62 & 1.00 & 6.0..... &  2.89 $\pm$  0.36 & 1.00 \\ 
7.0..... &  0.32 $\pm$  0.27 & 0.88 & 7.0..... &  4.07 $\pm$  0.51 & 1.00 & 7.0..... &  2.50 $\pm$  0.33 & 0.97 \\ 
8.0..... &  0.12 $\pm$  0.24 & 0.89 & 8.0..... &  3.81 $\pm$  0.47 & 1.00 & 8.0..... &  1.38 $\pm$  0.26 & 0.98 \\ 
9.0..... &  0.15 $\pm$  0.23 & 0.90 & 9.0..... &  2.81 $\pm$  0.41 & 0.99 & 9.0..... &  1.13 $\pm$  0.24 & 0.99 \\ 
10.0.... &  0.32 $\pm$ 0.24 & 0.90 & 10.0.... &  2.01 $\pm$ 0.39 & 0.81 & 10.0.... &  0.66 $\pm$ 0.21 & 0.98 \\ 
11.0.... & $-$0.15 $\pm$ 0.20 & 0.96 & 11.0.... &  1.53 $\pm$ 0.35 & 0.80 & 11.0.... &  0.89 $\pm$ 0.21 & 1.00 \\ 
12.0.... &  0.14 $\pm$ 0.22 & 0.95 & 12.0.... &  0.83 $\pm$ 0.31 & 0.78 & 12.0.... &  0.66 $\pm$ 0.20 & 0.98 \\ 
13.0.... &  0.21 $\pm$ 0.22 & 0.94 & 13.0.... &  0.23 $\pm$ 0.27 & 0.75 & 13.0.... &  0.41 $\pm$ 0.18 & 0.96 \\ 
14.0.... &  0.14 $\pm$ 0.21 & 0.91 & 14.0.... &  0.39 $\pm$ 0.28 & 0.75 & 14.0.... &  0.50 $\pm$ 0.18 & 0.99 \\ 
15.0.... &  0.05 $\pm$ 0.21 & 0.91 & 15.0.... &  0.33 $\pm$ 0.27 & 0.74 & 15.0.... &  0.10 $\pm$ 0.16 & 0.99 \\ 
16.0.... & $-$0.01 $\pm$ 0.20 & 0.93 & 16.0.... &  0.28 $\pm$ 0.27 & 0.74 & 16.0.... &  0.35 $\pm$ 0.17 & 0.98 \\ 
17.0.... & $-$0.16 $\pm$ 0.19 & 0.92 & 17.0.... &  0.19 $\pm$ 0.26 & 0.74 & 17.0.... &  0.34 $\pm$ 0.17 & 0.98 \\ 
18.0.... &  0.12 $\pm$ 0.21 & 0.69 & 18.0.... &  0.11 $\pm$ 0.27 & 0.54 & 18.0.... &  0.37 $\pm$ 0.17 & 0.85 \\ 
19.0.... &  0.13 $\pm$ 0.23 & 0.47 & 19.0.... &  0.05 $\pm$ 0.28 & 0.36 & 19.0.... &  0.15 $\pm$ 0.17 & 0.63 \\ 
20.0.... &  0.12 $\pm$ 0.24 & 0.36 & 20.0.... &  0.06 $\pm$ 0.29 & 0.26 & 20.0.... &  0.06 $\pm$ 0.18 & 0.45 \\ 
21.0.... &  0.34 $\pm$ 0.28 & 0.26 & 21.0.... &  0.06 $\pm$ 0.31 & 0.20 & 21.0.... &  0.17 $\pm$ 0.20 & 0.32 \\ 
22.0.... &  0.03 $\pm$ 0.26 & 0.18 & 22.0.... & $-$0.22 $\pm$ 0.29 & 0.14 & 22.0.... &  0.13 $\pm$ 0.22 & 0.23 \\ 
22.9.... & $-$0.13 $\pm$ 0.26 & 0.11 & 22.9.... &  0.13 $\pm$ 0.40 & 0.08 & 23.0.... & $-$0.10 $\pm$ 0.22 & 0.15 \\ 
23.9.... & $-$0.09 $\pm$ 0.36 & 0.04 & 23.9.... & $-$0.17 $\pm$ 0.43 & 0.04 & 23.9.... & $-$0.19 $\pm$ 0.25 & 0.09 \\ 
 & &  &  & & & 24.9.... &  0.04 $\pm$ 0.42 & 0.03 \\ 

\enddata
\tablecomments{A subtractive correction for contamination has been
applied to the data, resulting in negative surface densities in some
of the outer bins.}  \protect\label{table:Es profiles}
\end{deluxetable}

\begin{deluxetable}{lcccc}
\tablecaption{Coefficients from Fitting Radial Profile Data}
\tablewidth{370pt}
\tablehead{
\colhead{} & \multicolumn{2}{c}{deVaucouleurs Law} &
\multicolumn{2}{c}{Power Law}\\
\cline{2-3} \cline{4-5}\\
\colhead{Galaxy} & \colhead{a0}& \colhead{a1} & \colhead{a0}& \colhead{a1}}
\startdata
NGC~3379 & 2.33 $\pm$ 0.22 & $-$1.62 $\pm$ 0.15 & 0.83 $\pm$ 0.09 & $-$1.41 $\pm$ 0.13\\
NGC~4406 & 3.18 $\pm$ 0.09 & $-$1.58 $\pm$ 0.06 & 1.66 $\pm$ 0.04 & $-$1.24 $\pm$ 0.05\\
NGC~4594 & 3.76 $\pm$ 0.11 & $-$2.11 $\pm$ 0.08 & 1.87 $\pm$ 0.05 & $-$1.85 $\pm$ 0.07\\
\enddata
\protect\label{table:Es coefficients}
\end{deluxetable}

\begin{deluxetable}{llrlrlllr}
\tablecaption{Total Numbers and Specific Frequencies for Galaxies in the Early-type Sample}
\tablewidth{400pt}
\tablehead{\colhead{Name}
&\colhead{$M_V^T$} &\colhead{$N_{GC}$}&\colhead{$S_N$} &\colhead{Extent}
&\colhead{$T$}&\colhead{$T_{\rm blue}$}&\colhead{Prev $S_N$\tablenotemark{\dag}}
&\colhead{Ref}\\
\colhead{} & \colhead{} & \colhead{} & \colhead{} & \colhead{(kpc)} & \colhead{} & \colhead{} & \colhead{}
}
\startdata
NGC4472&$-$23.1 & 5900 & 3.6$\pm$0.6 & 100 & 4.2 & 2.6 &
4.5$\pm$1.3& 1\\
NGC4406&$-$22.3 & 2900 & 3.5$\pm$0.5 & 80 & 4.1 & 2.5 &
4.6$\pm$1.1 & 1\\
NGC3379&$-$20.9 & 270 & 1.2$\pm$0.3 & 30 & 1.4 & 1.0 &
1.1$\pm$0.6 & 1\\
NGC4594&$-$22.4 & 1900 & 2.1$\pm$0.3 & 50 & 3.2 &
2.0 & 2$\pm$1 & 2\\
\enddata \tablerefs{(1)\citet{harris91}; (2) \citet{bridges92}}
\tablenotetext{\dag}{The $S_N$ values given in this column were
calculated by combining $N_{GC}$ from the previous study with the
galaxy magnitude listed in column (2).}
\protect\label{table:Es S values}
\end{deluxetable}


\begin{thebibliography}{}
\bibitem[Ashman, Bird, \& Zepf (1994)]{abz94} Ashman, K.M., Bird,
C.M., \& Zepf, S.E. 1994, \aj, 108, 2348

\bibitem[Ashman \& Zepf (1992)]{az92} Ashman, K.M., \& Zepf, S.E. 1992,
\apj, 384, 50

\bibitem[Ashman \& Zepf (1998)]{az98} Ashman, K.M., \& Zepf, S.E. 1998,
Globular Cluster Systems (Cambridge: Cambridge University Press)

\bibitem[Baum (1955)]{baum55} Baum, W.A. 1955, \pasp, 67, 328

\bibitem[Beasley et al.\ (2002)]{beasley02} Beasley, M.A., Baugh,
C.M., Forbes, D.A., Sharples, R.M., \& Frenk, C.S. 2002, \mnras, 333, 383

\bibitem[Bridges \& Hanes (1992)]{bridges92} Bridges, T.J. \& Hanes,
D.A. 1992, \aj, 103, 800

\bibitem[Cohen (1988)]{cohen88} Cohen, J.G. 1988, \aj, 95, 682

\bibitem[C\^ot\'e et al.\ (1998)]{cote98} C\^ot\'e, P., Marzke, R.O., \& West,
M.J. 1998, \apj, 501, 554

\bibitem[de Vaucouleurs et al.\ (1991)]{devauc91} de~Vaucouleurs, G.,
de~Vaucouleurs, A., Corwin Jr., H.G., Buta, R.J., Paturel, G., \&
Fouque, P. 1991, Third Reference Catalogue of Bright Galaxies (New
York: Springer)


\bibitem[Dressler et al.\ (1987)]{dressler87} Dressler, A.,
Lynden-Bell, D., Burstein, D., Davies, R.L., Faber, S.M., Terlevich,
R.J., \& Wegner, G. 1987, \apj, 313, 42

\bibitem[Faber et al.\ (1987)]{faber87} Faber, S.M., Dressler, A.,
Davies, R.L., Burstein, D., \& Lynden-Bell, D. 1987, in Nearly Normal
Galaxies: From the Planck Time to the Present (New York:
Springer-Verlag)

\bibitem[Forbes et al.\ (1997a)]{fbg97} Forbes, D.A., Brodie, J.P., \&
Grillmair, C.J. 1997a, \aj, 113, 1652

\bibitem[Forbes et al.\ (1997b)]{forbes97} Forbes, D.A., Grillmair,
C.J., \& Smith, R.C. 1997b, \aj, 113, 1648

\bibitem[Gebhardt \& Kissler-Patig (1999)]{geb99} Gebhardt, K. \&
Kissler-Patig, M. 1999, \aj, 118, 1526

\bibitem[Geisler, Lee, \& Kim (1996)]{geisleretal96} Geisler, D., Lee,
M.G., \& Kim, E. 1996, \aj, 111, 1529

\bibitem[Harris (1991)]{harris91} Harris, W.E. 1991, \araa, 29, 543


\bibitem[Harris \& van den Bergh (1981)]{hvdb81} Harris, W.E. \& van
den Bergh, S. 1981, \aj, 86, 1627

\bibitem[Kormendy \& Djorgovski (1989)]{korm89} Kormendy, J. \&
Djorgovski, S. 1989, \araa, 27, 235

\bibitem[Kundu \& Whitmore (2001)]{kw01} Kundu, A. \& Whitmore,
B.C. 2001, \aj, 121, 2950

\bibitem[Kundu et al.\ (1999)]{kundu99} Kundu, A., Whitmore, B.C.,
Sparks, W.B., \& Macchetto, F.D. 1999, \apj, 513, 733

\bibitem[Landolt (1992)]{land92} Landolt, A.U. 1992, \aj, 104, 340

\bibitem[Larsen et al.\ (2001)]{larsen01} Larsen, S.S., Brodie, J.P.,
Huchra, J.P., Forbes, D.A., \& Grillmair, C.J. 2001, \aj, 121, 2974

\bibitem[Mendez et al.\ (2000)]{mendez00} Mendez, R.A., Platais, I.,
Girard, T.M., Kozhurina-Platais, V., \& van Altena, W.F. 2000, \aj,
119, 813

\bibitem[Mendez \& van Altena (1996)]{mendez96} Mendez, R.A. \& van
Altena, W.F. 1996, \aj, 112, 655

\bibitem[Pahre et al.\ (1995)]{pahre95} Pahre, M.A., Djorgovski, S.G.,
\& de~Carvalho, R.R. 1995, \apj, 453, L17

\bibitem[Rhode \& Zepf (2001)]{rhode01} Rhode, K.L. \& Zepf,
S.E. 2001, \aj, 121, 210

\bibitem[Rhode \& Zepf (2003)]{rhode03} Rhode, K.L. \& Zepf,
S.E. 2003, \aj, in press

\bibitem[Sandage (1961)]{sandage61} Sandage, A. 1961, ``The Hubble
Atlas of Galaxies'' (Washington, D.C.: Carnegie Institution of
Washington)

\bibitem[Santos (2003)]{santos03} Santos, M.R. 2003, in Extragalactic
Globular Cluster Systems, ed. M. Kissler-Patig (New York:
Springer-Verlag)

\bibitem[Schlegel, Finkbeiner, \& Davis (1998)]{schlegel98} Schlegel,
D.J., Finkbeiner, D.P., \& Davis, M. 1998, \apj, 500, 525

\bibitem[Tonry et al.\ (2001)]{tonry01} Tonry, J.L., Blakeslee, J.P.,
Ajhar, E.A., Fletcher, A.B., Luppino, G.A., Metzger, M.R., \& Moore,
C.B. 2001, \apj, 546, 681

\bibitem[Whitmore \& Schweizer (1995)]{whitschweiz95} Whitmore, B.C.
\& Schweizer, F. 1995, \aj, 109, 960

\bibitem[Whitmore et al.\ (1993)]{whitmore93} Whitmore,
B.C., Schweizer, F., Leitherer, C., Borne, K., \& Robert, C. 1993,
\aj, 106, 1354

\bibitem[Whitmore et al.\ (1995)]{whitmore95} Whitmore, B.C., Sparks,
W.B., Lucas, R.A., Macchetto, F.D., \& Biretta, J.A. 1995, \apj, 454,
L73

\bibitem[Zepf \& Ashman (1993)]{za93} Zepf, S.E., \& Ashman, K.M. 1993,
\mnras, 264, 611

\bibitem[Zepf \& Silk (1996)]{zs96} Zepf, S.E. \& Silk, J. 1996, \apj,
466, 114

\end{thebibliography}
\end{document}